\documentclass[%
reprint,
amsmath,amssymb,prx,
aps, longbibliography, groupedaddress,superscriptaddress
]{revtex4-1}

\usepackage{graphicx}
\usepackage{dcolumn}
\usepackage{bm}
\usepackage{braket}
\usepackage{mathtools}
\usepackage{bbm}
\usepackage{comment}
\usepackage{makecell}
\usepackage{afterpage}
\usepackage{amsthm}

\newcommand\blankpage{%
	\null
	\thispagestyle{empty}%
	\addtocounter{page}{-1}%
	\newpage}

\usepackage{hyperref}
\hypersetup{
	colorlinks=true,
	linkcolor=blue,
	citecolor=blue,
	filecolor=magenta,
	urlcolor=cyan
}

\def\iu{{\rm i}}

\DeclareMathOperator{\Tr}{Tr}
\DeclareMathOperator{\Var}{Var}
\DeclareMathOperator{\Cov}{Cov}
\def\dif{{\rm d}}
\def\pstring{\boldsymbol{\sigma}}

\newtheorem*{theorem*}{Theorem}
\newtheorem*{lemma*}{Lemma}
\newtheorem*{claim*}{Claim}

\makeatletter
\DeclareFontFamily{OMX}{MnSymbolE}{}
\DeclareSymbolFont{MnLargeSymbols}{OMX}{MnSymbolE}{m}{n}
\SetSymbolFont{MnLargeSymbols}{bold}{OMX}{MnSymbolE}{b}{n}
\DeclareFontShape{OMX}{MnSymbolE}{m}{n}{
	<-6>  MnSymbolE5
	<6-7>  MnSymbolE6
	<7-8>  MnSymbolE7
	<8-9>  MnSymbolE8
	<9-10> MnSymbolE9
	<10-12> MnSymbolE10
	<12->   MnSymbolE12
}{}
\DeclareFontShape{OMX}{MnSymbolE}{b}{n}{
	<-6>  MnSymbolE-Bold5
	<6-7>  MnSymbolE-Bold6
	<7-8>  MnSymbolE-Bold7
	<8-9>  MnSymbolE-Bold8
	<9-10> MnSymbolE-Bold9
	<10-12> MnSymbolE-Bold10
	<12->   MnSymbolE-Bold12
}{}

\let\llangle\@undefined
\let\rrangle\@undefined
\DeclareMathDelimiter{\llangle}{\mathopen}%
{MnLargeSymbols}{'164}{MnLargeSymbols}{'164}
\DeclareMathDelimiter{\rrangle}{\mathclose}%
{MnLargeSymbols}{'171}{MnLargeSymbols}{'171}
\makeatother

\begin{document}
	
	\def\papertitle{{Quantifying information scrambling via Classical Shadow Tomography on Programmable Quantum Simulators}}
	
	\title{\papertitle}
	
	\def\oxf{{Rudolf Peierls Centre for Theoretical Physics, Clarendon Laboratory, Oxford University, Parks Road, Oxford OX1 3PU, United Kingdom}}
	\def\berk{{Department of Physics, University of California, Berkeley, California 94720, USA}}
	
	\author{Max McGinley}
	\affiliation{\oxf}
	\author{Sebastian Leontica}
	\affiliation{\oxf}
	\author{Samuel J.~Garratt}
	\affiliation{\oxf}
	\affiliation{\berk}
	\author{Jovan Jovanovic}
	\affiliation{\oxf}
	\author{Steven H.~Simon}
	\affiliation{\oxf}
	
	\def\authorlist{Max McGinley, Sebastian Leontica, Samuel J.~Garratt, Jovan Jovanovic, and Steven H.~Simon}

	\date{\today}
	
	\begin{abstract}
	We develop techniques to probe the dynamics of quantum information, and implement them experimentally on an IBM superconducting quantum processor. Our protocols adapt shadow tomography for the study of time evolution channels rather than of quantum states, and rely only on single-qubit operations and measurements.
	We identify two unambiguous signatures of quantum information scrambling, neither of which can be mimicked by dissipative processes, and relate these to many-body teleportation.
	By realizing quantum chaotic dynamics in experiment, we measure both signatures, and support our results with numerical simulations of the quantum system. We additionally investigate operator growth under this dynamics, and observe behaviour characteristic of quantum chaos. As our methods require only a single quantum state at a time, they can be readily applied on a wide variety of quantum simulators.
	\end{abstract}
	
	\maketitle
	
	\section{Introduction}
	
	
	Scrambling is fundamental to our current understanding of many-body quantum dynamics in fields ranging from thermalization and chaos \cite{Deutsch1991,Srednicki1994,Tasaki1998,Rigol2008} to black holes \cite{Hayden2007,Sekino2008,Shenker2014}. This is the process by which initially local information, such as charge imbalance in a solid, becomes hidden in increasingly non-local degrees of freedom under unitary time evolution. Scrambling accounts for both the fate of information falling into black holes \cite{Hawking1976, Hayden2007}, as well as the apparent paradox of equilibration under unitary dynamics: Information about the initial state is not truly lost, but rather becomes inaccessible when one can only measure local observables, as is the case in traditional experimental settings. 
	
	Today, the experimental settings we have access to offer a much higher degree of control and programmability than those that were available when these questions were first addressed. New kinds of quantum devices can be constructed by assembling qubits that are individually addressable, such as those made from trapped ions \cite{Benhelm2008,Nigg2014,Zhang2017,Friis2018}, superconducting circuits \cite{Barends2014,Kelly2015,Ofek2016,Wendin2017,Mi2022}, or Rydberg atoms \cite{Weimer2010,Barreiro2011,Barredo2016,Endres2016,Bernien2017,Ebadi2021}. Such noisy intermediate scale quantum (NISQ) devices \cite{Preskill2018} allow a wider range of interactions to be synthesised, and, crucially, permit measurements of highly non-local observables, making the distinction between non-unitary information loss and unitary information scrambling more than a purely academic one. As well as providing further motivation for theoretical work on quantum chaos and scrambling, these technological developments open the door to complementary experimental studies, which promise to be of increasing utility as the size and complexity of the systems continue to grow beyond what can be simulated classically \cite{Arute2019,Zhong2020}.
	
	A variety of experimental protocols to probe quantum chaos have already been put forward and implemented, with early approaches based on measuring the growth of quantum entanglement. For example, if two copies of the system can be prepared simultaneously, then certain quantifiers of entanglement can be extracted from joint measurements on the two copies \cite{Ekert2002,Moura2004,Daley2012,Pichler2013,Islam2015}. More recently, focus has shifted towards probing scrambling rather than entanglement growth, primarily via so-called out-of-time-order correlators (OTOCs) \cite{Larkin1969,Shenker2014,Kitaev2014,Roberts2015,Aleiner2016}, which can be measured when the dynamics can be time-reversed \cite{Li2017,Garttner2017,Wei2018,Joshi2020,Mi2021}. However, the link between OTOC decay and scrambling is predicated on the assumption that the dynamics is unitary \cite{Yoshida2019} --- this is invariably not the case in NISQ devices, which are by definition noisy. Moreover, with system sizes being somewhat limited at present, protocols that are qubit-efficient (i.e.~not requiring multiple copies of the system at once) will be required to make progress in the near term.

	Emphasising its practical implementation in an IBM superconducting quantum computer, in this work we show how scrambling can be quantified in NISQ devices using only single-qubit manipulations and individual copies of a quantum state at a time. To achieve this, we first generalise the technique of shadow tomography \cite{Huang2020} to study dynamics. We then prove that certain well-established physical quantities are (i) accessible using this technique, and (ii) provide unambiguous signatures of scrambling. Crucially, the signatures that we identify remain meaningful even when the system's dynamics is non-unitary; this allows us to verifiably detect scrambling on a real noisy quantum device.
	
	The quantities that we identify satisfying the above two criteria are related to operator-space entanglement (OE), also known as entanglement in time \cite{Zanardi2000,Zanardi2001,Hosur2016, Lensky2019}. While entanglement quantifies quantum correlations between degrees of freedom at one instant in time, OE pertains to correlations that are conveyed across time, which is of direct relevance to scrambling. This has proved to be an extremely useful tool in analytical and numerical studies of chaotic quantum dynamics \cite{Hosur2016,Zhou2017,Dubail2017,Iyoda2018,Pal2018,Nie2019,Schnaack2019,Bertini2020,Styliaris2021}, allowing one to construct measures of chaos in a dynamical, manifestly state-independent way.
	
	Here we establish a link between OE and the ability of a system to transmit information from one qubit to another via a process known as many-body teleportation, or Hayden-Preskill teleportation after the authors of Ref.~\cite{Hayden2007}. This process was originally considered in the context of the black hole information paradox \cite{Hawking1976}, and is now a central part of the theory of scrambling. We put forward two OE-based quantities [Eqs.~(\ref{eq:RenyiMutual}, \ref{eq:RRatio})], and show that each can be related to the fidelity of Hayden-Preskill teleportation. In particular, we argue that both quantities have a threshold value which when exceeded gives a guarantee that the quantum communication capacity from one qubit to another is non-zero, i.e.~quantum states can be reliably transmitted at a finite rate using the quantum system as a communication channel, even when dynamics is non-unitary.
	
	Beyond establishing these quantities as meaningful measures of scrambling, we demonstrate their practical utility by showing that both are directly measurable in experiment. The scheme we introduce allows one to measure the necessary information-theoretic quantities with minimal experimental overhead. This is made possible by extending ideas originally developed to measure entanglement in an instantaneous state. In that context, it has been demonstrated that measurements of the state in randomly selected bases can be used to extract certain entanglement measures \cite{Elben2018, Brydges2019, Huang2020}, without requiring joint access to multiple copies of the state per experiment. To generalize from state entanglement to OE, we propose to prepare initial states in random bases, which are then time evolved under the dynamics of interest, before being measured in random bases (see Fig.~\ref{fig:Protocol}). By post-processing the classical data generated by this sequence of operations in a way analogous to that proposed in Ref.~\cite{Huang2020}, we are able to construct estimators of the quantities in question. We do so explicitly using data from an IBM quantum computer, giving us access to spatially-resolved measures of information delocalization, revealing the light-cone structure in the system's dynamics.

	In addition to these probes of many-body teleportation, our protocol can be used to obtain a fine-grained description of operator spreading \cite{Roberts2015, Nahum2018,vonKeyserlingk2018,Khemani2018}. Specifically, shadow tomography of the dynamics gives us access to certain combinations of the operator spreading coefficients studied in Ref.~\cite{vonKeyserlingk2018}, which gives a complementary perspective on scrambling.
	
	Other quantities related to operator entanglement, namely out-of-time-order correlators (OTOCs) \cite{Larkin1969,Shenker2014,Kitaev2014,Roberts2015,Aleiner2016}, have been measured in previous experiments \cite{Li2017,Garttner2017,Wei2018,Joshi2020,Mi2021}, and indeed are in principle measurable using shadow tomography and related methods \cite{Vermersch2019,Garcia2021}. However, these cannot be used as an unambiguous diagnostic of scrambling, since dissipation and miscalibrations can give rise to the same signal as that of a true scrambler \cite{Yoshida2019}. In contrast, the quantities (\ref{eq:RenyiMutual}, \ref{eq:RRatio}) measured here constitute a positive, verifiable signature of scrambling, which cannot be mimicked by noise. We note that related signatures of teleportation have been observed before using multiple copies of the system evolving in a coordinated fashion \cite{Landsman2019, Blok2021}. A key innovation in our work is to quantify the fidelity of teleportation without actually performing teleportation. As a consequence our method can probe scrambling with half as many qubits, and without needing to match the time evolution between two separate systems, which may not be possible when the dynamics is not known \textit{a priori}.

	As well as superconducting qubits, the protocol we use here is implementable using presently available techniques in a variety of other platforms including those based on Rydberg atom arrays \cite{Weimer2010,Barredo2016,Endres2016,Bernien2017}, trapped ions \cite{Zhang2017,Friis2018}, and photonics \cite{Peruzzo2014,Carolan2015,Flamini2018,Zhong2020}. We compare the protocol to previous approaches used to diagnose scrambling, and discuss the tradeoffs between sample efficiency, verifiability, and the required degree of experimental control.
	
	This paper is organised as follows. In Section \ref{sec:OpEntHP}, we introduce the concept of operator-space entanglement, as well as the Hayden-Preskill protocol for many-body teleportation \cite{Hayden2007}, and describe how the two are related. We then introduce the key quantities (\ref{eq:RenyiMutual}, \ref{eq:RRatio}) that we will use to quantify Hayden-Preskill teleportation in Section \ref{sec:RenyiScrabling}, as well as showing how OE allows one to track the growth of operators under Heisenberg time evolution. Section \ref{sec:Protocol} describes our shadow tomographic protocol that can be used to estimate the above quantities. Results from implementing this protocol on an IBM superconducting quantum processor are given in Section \ref{sec:IBM}. We discuss our results and present our conclusions in Section \ref{sec:Discussion}.

	\section{Probing scrambling using operator-space entanglement}
	
	\subsection{Operator-space entanglement and the Hayden-Preskill protocol \label{sec:OpEntHP}}
	
	The evolution of a quantum system $Q$ with Hilbert space $\mathcal{H}_Q$ from time $0$ to time $t$ can be described by a channel $\mathcal{N}_t$, such that the density matrix evolves as $\rho^Q(t) = \mathcal{N}_t[\rho^Q(0)]$. The usual notion of entanglement in a state can be generalized to channels, which is known as operator-space entanglement. Formally, this is done by reinterpreting $\mathcal{N}_t$ as a state on a doubled Hilbert space  \cite{Jamiolkowski1972,Choi1975}, on which conventional entanglement measures can be defined. This is perhaps most simply understood when the dynamics is unitary $\mathcal{N}_t[\rho^Q] = U_t \rho^Q U_t^\dagger$, as detailed in Ref.~\cite{Hosur2016}. Fixing a basis of product states $\{\ket{a}\}$ for $Q$, a pure doubled state (living in `operator space') is constructed as $\ket{U_t}_{\rm op} = |\mathcal{H}_Q|^{-1/2} \sum_{ab} \braket{b|U_t|a} \ket{a}_{\rm in} \otimes \ket{b}_{\rm out}$, where $|\mathcal{H}_Q|$ is the Hilbert space dimension, and the `in' and `out' labels refer to the inputs and outputs of the unitary. In words, the components of $\ket{U_t}_{\rm op}$ are the $|\mathcal{H}_Q|^2$ matrix elements of the unitary $U_t$. From here onwards we specialize to $N$-qubit systems, so $|\mathcal{H}_Q| = 2^{N}$.
	
\begin{figure}
		\includegraphics[width=246pt]{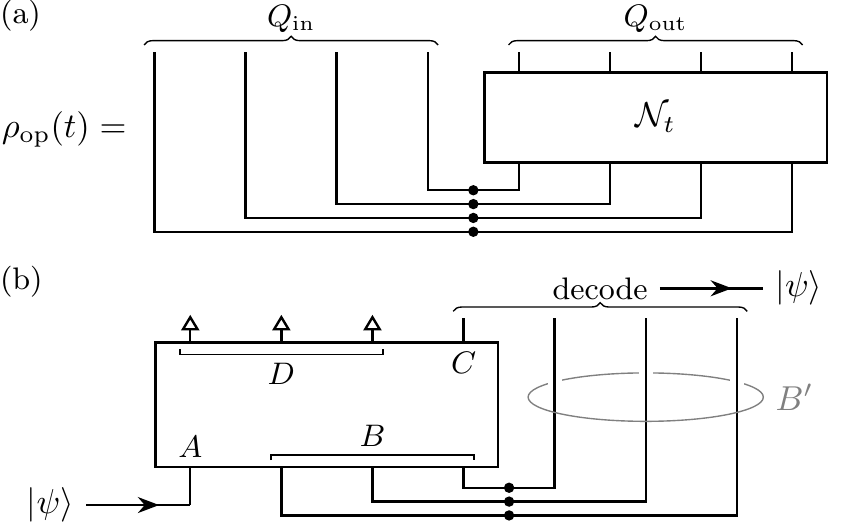}
		\caption{(a) Representation of the operator state $\rho_{\rm op}(t)$ [Eq.~\eqref{eq:RhoOp}]. Each qubit in $Q_{\rm out}$ is prepared in a maximally entangled state (black dots) with the corresponding qubit $Q_{\rm in}$, before being time evolved under the channel $\mathcal{N}_t$. (b) Illustration of the Hayden-Preskill protocol \cite{Hayden2007}. An unknown quantum state $\ket{\psi}$ is used as an input to a small subregion $A$, while the remaining qubits ($B$) are prepared in a maximally entangled state with a set of ancillas $B'$ (circled). If the channel is perfectly scrambling then $\ket{\psi}$ can be reconstructed using the ancillas combined with a subset of output qubits $C$ of the same size as $A$, regardless of which qubits are in $C$ (qubits in $D$ are discarded). Formally, the final state of the ancillas combined with the outputs $C$ depends on the input state to $A$ through the channel $\mathcal{N}^{A \rightarrow B'C}_t$ (see main text).}
		\label{fig:HaydenPreskill}
	\end{figure}

	This construction has an alternative interpretation: $\ket{U_t}_{\rm op}$ is the state that results from evolving a maximally entangled state $\ket{\Phi} = 2^{-N/2}\sum_a \ket{a}_{\rm in} \otimes \ket{a}_{\rm out}$ under the unitary $\mathbb{I}_{\rm in} \otimes U_t$, i.e.~one half of the maximally entangled pair is evolved under $U_t$. This also makes it clear how to generalize to non-unitary evolutions: $\ket{U_t}_{\rm op}$ is replaced by a mixed state
	\begin{align}
	\rho_{\rm op}(t) = ({\rm id}_{\rm in} \otimes \mathcal{N}_t)[\ket{\Phi}\bra{\Phi}].
	\label{eq:RhoOp}
	\end{align}
	This construction is illustrated in Fig.~\ref{fig:HaydenPreskill}(a). We use the more generally applicable density matrix $\rho_{\rm op}(t)$, rather than the pure state $\ket{U_t}_{\rm op}$, in the following. Note that correlation functions with respect to the doubled state $\Tr[(O_{\rm in} \otimes O_{\rm out}) \rho_{\rm op}(t)]$ map to infinite-temperature two-time correlation functions $2^{-N}\Tr[O_{\rm in}^T O_{\rm out}(t)]$ (where time evolution of operators in the Heisenberg picture is given by $O_{\rm out}(t) = \mathcal{N}_t^\dagger[O_{\rm out}]$, and the transpose is taken with respect to the basis $\{\ket{a}\}$).
	
	Evidently, at $t = 0$ ($\mathcal{N}_{t=0} = \text{id}$) a given input qubit with index $j_{\rm in}$ is maximally entangled with the corresponding output qubit $j_{\rm out} = j_{\rm in}$ only. This reflects the trivial observation that information is perfectly transmitted from $j_{\rm in}$ to $j_{\rm out} = j_{\rm in}$ under $\mathcal{N}_{t = 0}$. If $\mathcal{N}_t$ exhibits scrambling, then we expect that locally encoded information will begin to spread out as the output qubits evolve, such that $j_{\rm in}$ becomes entangled with many other output qubits. At late times, one will no longer be able to extract these correlations from any small output region $C$; instead, the information about the initial state of a given qubit will be encoded across many output qubits.
	
	This intuition can be quantified in terms of particular measures of operator-space entanglement. These are constructed by evaluating familiar quantities associated with state entanglement on $\rho_{\rm op}(t)$. In the doubled space, one can divide the input qubits into $A$ and its complement $B$, and the outputs into $C$ and its complement $D$. ($A$ and $C$ need not correspond to the same physical qubits.) Reduced density matrices can then be formed, e.g.~$\rho^{AC}(t) = \Tr_{B \cup D} \rho_{\rm op}(t)$. Two important information-theoretic quantities are the von Neumann entanglement entropy $S(AC) = -\Tr \rho^{AC}(t) \log \rho^{AC}(t)$, and the mutual information $I(A:C) = S(A) + S(C) - S(AC)$ (all logarithms are base-2, and we leave the $t$-dependence of entropies and mutual information implicit). The mutual information quantifies the degree to which the initial state of qubits in $A$ is correlated with the final state of qubits in $C$ (this includes both classical and quantum correlations). Indeed, $I(A:C)$ is closely related to the capacity of the channel for classical communication from a sender $A$ to a receiver $C$ \cite{Schumacher1997,Holevo1998}.
	
	Given that the reduced density matrices $\rho^{AC}(t)$ will typically be highly mixed, it is also useful to examine quantities that have been devised to probe mixed state entanglement. The logarithmic negativity $E_{A:C}\coloneqq \log \Tr |\rho^{AC}(t)^{T_A}|$ (where $T_A$ denotes a partial transpose on $A$ and $| O | \coloneqq \sqrt{O^\dagger O}$ for operators $O$) is useful for this purpose: when applied to a bipartite state it can be used to bound the distillable entanglement between $A$ and $C$ \cite{Vidal2002,Plenio2005}, which unlike mutual information excludes classical correlations. Here we will consider the operator-space generalization of negativities, which have been connected to scrambling in the context of random unitary circuits and holographic channels \cite{Kudler2020}.

	As argued by the authors of Ref.~\cite{Hosur2016}, for unitary chaotic channels the correlations between regions $A, C$ of size $\mathcal{O}(1)$ will be small, whereas $I(A:CD)$ will be maximal, indicating that the input state $A$ can only be reconstructed if one has access to all the outputs $CD$. They propose the tripartite information $I_3(A:C:D) = I(A:C) + I(A:D) - I(A:CD)$ as a diagnostic of scrambling (for scramblers $I_3$ is large and negative), illustrating one way in which operator-space entanglement measures can be used to detect scrambling.
	
	A complementary way to diagnose scrambling is to quantify correlations between $A$ and $BC$ that are present in $\rho_{\rm op}(t)$, where again $A, C$ are of size $\mathcal{O}(1)$. This approach is related to the Hayden-Preskill teleportation problem \cite{Hayden2007} -- a thought experiment that was initially devised to understand the fate of information in black holes. There, one asks if it is possible to recover the initial state of a small set of qubits $A$ using a set of ancillas $B'$ that are initially maximally entangled with $B$, combined with a subset of output qubits $C$, see Fig.~\ref{fig:HaydenPreskill}(b). If $\mathcal{N}_t$ is scrambling, then the initial state of $A$ becomes non-locally encoded across the entire system. When this occurs, teleportation can be achieved (i.e.~the initial state of $A$ can be recovered from $B'C$) regardless of which qubits are chosen in $C$, as long as $|C| \geq |A|$ \cite{Landsman2019}. 
	
	Intuitively, we expect that for teleportation to be successful, there must be strong correlations between $A$ and $BC$ in the state $\rho_{\rm op}(t)$. This can in principle be diagnosed using the quantities introduced above, namely $I(A:BC)$ and $E_{ A:BC}$. More formally, we can capture the dependence of the final state of $B'C$ on the initial state $A$ using the channel $\mathcal{N}^{A \rightarrow B'C}_t[\rho^A]  = \Tr_D[(\mathcal{N}_t \otimes \text{id}_{B'})[\rho^A \otimes \Phi_{BB'}]]$. The fidelity of teleportation in the Hayden-Preskill protocol is then determined by the potential for information transmission through $\mathcal{N}^{A \rightarrow B'C}_t$, which can be quantified in an information-theoretic way using an appropriate channel capacity \cite{Nielsen2010}. As an example, the classical capacity of $\mathcal{N}^{A \rightarrow B'C}_t$ is closely related to $I(A:BC)$ \cite{Schumacher1997,Holevo1998}. Similarly, the quantum channel capacity (the maximum rate at which quantum states can be reliably transmitted using multiple applications of the channel) can be bounded by $E_{A:BC}$ \cite{Pisarczyk2019}. This illustrates the connection between information transmission in the Hayden-Preskill protocol and the degree of correlations between $A$ and $BC$ in the operator state $\rho_{\rm op}(t)$.
	
	The experiment of Ref.~\cite{Landsman2019} provided an explicit demonstration of scrambling by executing a particular decoding procedure for the Hayden-Preskill protocol. This requires one to construct a doubled state, and manipulate the ancillas $B'$. In contrast, in this paper we will employ a different approach, where we quantify the correlations between $A$ and $BC$ without ever performing the teleportation explicitly, and relate these to properties of $\mathcal{N}^{A \rightarrow B'C}_t$. This avoids us having to construct a doubled state or execute a decoding procedure.
	
	\subsection{R{\'e}nyi measures of scrambling and operator growth \label{sec:RenyiScrabling}}
	
	While the von Neumann entropy and quantities derived thereof have strong information-theoretic significance, they are not directly measurable in experiments without recourse to full tomography of $\rho_{\rm op}(t)$, which is computationally expensive \cite{Gross2010}. This is due to the need to take the operator logarithm of $\rho$.
	Instead, one can generalize to R{\'e}nyi entropies $S^{(m)}(AC) \coloneqq (1-m)^{-1}\log \Tr([\rho^{AC}(t)]^m)$ ($m = 2,3, \ldots$), which unlike $S({AC})$ only depend on integer moments of the density matrix, and hence can be computed in terms of $m$th moments of correlation functions of $\rho^{AC}(t)$.
	This observation forms the basis of a number of protocols which use randomized measurements to extract the R{\'e}nyi entropies of an instantaneous state \cite{Elben2018,Huang2020}, as well as integer moments of the density matrix after partial transposition \cite{Elben2020}. Later, we will employ similar arguments to show that the analogous quantities in operator space can also be directly measured. Before doing so, we first discuss how these quantities can be used to probe quantum chaotic dynamics and information scrambling, making use of the insight described in the previous section.\\
	
	We have argued how $I(A:BC)$ can be related to the fidelity of the Hayden-Preskill protocol. A natural generalization of $I(A:BC)$ that is constructed in terms of integer moments of $\rho_{\rm op}$ is the R{\'e}nyi mutual information
	\begin{align}
	I^{(m)}(A:BC) \coloneqq S^{(m)}(A) + S^{(m)}(BC) - S^{(m)}(ABC).
	\label{eq:RenyiMutual}
	\end{align}
	When evaluated on arbitrary states this simple generalization of the mutual information does not satisfy all the same properties as $I(A:BC)$, including non-negativity \cite{Wilde2014,Berta2015,Scalet2021}. However, in Appendix \ref{app:RenyiMI} we show that when evaluated on operator-states \eqref{eq:RhoOp} (for which the reduced density matrix on $A$ is maximally mixed), $I^{(m)}(A:BC)$ is non-negative \cite{Lensky2019}, and equal to zero if and only if $A$ and $BC$ are uncorrelated, as one would desire for any measure of correlation. Additionally, for $m = 2$ the R{\'e}nyi mutual information is related to the recovery fidelity $F$ for the decoding protocol used in Ref.~\cite{Landsman2019} by $F = 2^{I^{(2)}(A:BC) - 2|A|}$ \cite{Yoshida2019}, and can also be expressed in terms of particular sums of two-point correlation functions or OTOCs \cite{Hosur2016, Lensky2019}.
	
	Given the above, we expect that the quantity \eqref{eq:RenyiMutual} will be sensitive to the temporal correlations that are conveyed by channels that exhibit scrambling. Moreover, while mutual information captures classical and quantum correlations on an equal footing, one can still use $I^{(m)}(A:BC)$ to detect the transmission of purely quantum information. Specifically, we argue that the channel $\mathcal{N}^{A \rightarrow B'C}_t$, which describes the Hayden-Preskill setup, must have a non-zero quantum communication capacity if $I^{(m)}(A:BC)$ exceeds the threshold value of $|A|$, which is the maximum value that can be obtained in a classical system. The full proof of this statement is given in Appendix \ref{app:RenyiMI}. In brief, we show that violation of the classical limit can only occur if there is entanglement between $A$ and $BC$ in the operator state $\rho_{\rm op}(t)$. Given multiple uses of the channel, one can distil this entanglement into EPR pairs, which can then be used for noiseless quantum communication. This confirms that $\mathcal{N}^{A \rightarrow B'C}_t$ can in principle be used to reliably transmit quantum information, and thus the quantum capacity is non-zero. Note that the converse is not necessarily true, i.e.~there exist channels for which the quantum capacity is non-zero, but $I^{(m)}(A:BC) \leq |A|$.

	We can also consider quantities related to negativity that only involve integer moments of the density matrix. Let us first define moments of the partially transposed operator state $p_{m, X:Y} \coloneqq \Tr[(\rho^{XY}(t)^{T_X})^m]$, where $X$ and $Y$ are non-overlapping sets of input and output qubits, and again $T_X$ denotes a partial transpose on $X$. We will consider the quantity
	\begin{align}
	R_{A:BC} \coloneqq \frac{p_{2,A:BC}^2}{p_{3,A:BC}}.
	\label{eq:RRatio}
	\end{align}
	 This particular ratio was proposed as a measure of mixed state entanglement in Ref.~\cite{Elben2020}, where it was shown that bipartite states $\rho^{AB}$ satisfying $R_{A:B} > 1$ must be entangled. In Appendix \ref{app:RenyiMI}, we argue that $R_{A:BC} > 1$ is a sufficient (but not necessary) condition for the quantum communication capacity of $\mathcal{N}^{A \rightarrow B'C}_t$ to be non-zero, provided that $A$ is a single qubit (which is the case throughout this paper). \\
	
	The above arguments demonstrate how the R{\'e}nyi generalizations of mutual information and negativity can be related to the Hayden-Preskill teleportation fidelity. A complementary way to probe aspects of chaos in quantum dynamics is to consider the time evolution of operators in the Heisenberg picture $O(t) = \mathcal{N}^\dagger_t[O]$ \cite{Nahum2018,vonKeyserlingk2018}. Operator-space R{\'e}nyi entropies for $m = 2$ (equivalently, operator-space purities $\Tr[\rho^{AC}(t)^2] \equiv 2^{-S^{(2)}({AC})}$) can be related to the structure of operator growth. To see this, let us use Pauli strings $\pstring^\mu = \bigotimes_j \sigma_j^{\mu_j}$ as a basis of operators, where $\mu = (\mu_1, \ldots, \mu_N)$  and $\mu_j \in \{I, X, Y, Z\}$. Adopting the notation of Ref.~\cite{vonKeyserlingk2018}, operator spreading coefficients $c^{\mu \nu}(t)$ can then be defined via an expansion of time-evolved Pauli strings $\pstring^\mu(t) = \mathcal{N}_t^\dagger[\pstring^\mu]$, namely $\pstring^\mu(t) = \sum_{\nu} c^{\mu \nu}(t) \pstring^\nu$. It is straightforward to show that operator-space purity can be expressed succinctly in terms of operator spreading coefficients as
	\begin{align}
	\Tr[\rho^{AC}(t)^2] = \frac{1}{2^{|A|+|C|}}\sum_{\nu \in A} \sum_{\mu \in C} |c^{\mu \nu}(t)|^2,
	\label{eq:PurityOpspread}
	\end{align}
	where the sums are over Pauli strings $\nu$ and $\mu$ that act as identity on qubits outside of $A$ and $C$, respectively. In words, we identify operator-space purity as the norm of the part of the evolved operator $\pstring^\mu(t)$ that has support on $A$, averaged over all initial operators $\pstring^\mu$ with support on $C$.
	
	Eq.~\eqref{eq:PurityOpspread} clarifies how operator purities encode the spatial structure of operator spreading. One concise way to represent this information is in terms of the $k$-locality of the evolved operator $\pstring^\mu(t)$, i.e.~one can ask what proportion of the Pauli strings that make up $\pstring^\mu(t)$ act non-trivially on at most $k$ qubits. Intuitively, local operators with support on a small number of qubits will grow under chaotic time evolution, leading to more weight on operators that have a wider support. This contrasts with integrable systems, where $\pstring^\mu(t)$ spreads out in space without becoming more complex in terms of $k$-locality.
	
	A natural way to measure $k$-locality of the evolved operator $\pstring^\mu(t)$ is to compute the norm of the part of the operator that is made up of Pauli strings acting on exactly $k$ qubits
	\begin{align}
		D_k^\mu(t) &\coloneqq \sum_{\nu : |\pstring^\nu| = k} |c^{\mu \nu}(t)|^2.
	\end{align}
	where we use $|\pstring^\nu|$ to denote the number of non-identity factors in the string $\pstring^\nu$. If one takes an average of $D_k^\mu(t)$ over all non-identity Pauli strings $\mu$ with support in some region $C$, the resulting quantity can be expressed in terms of operator purities
	\begin{align}
	D_k^C(t) &\coloneqq \frac{1}{2^{|C|} - 1}\sum_{\mu \in C;\, \mu \neq I^{\times N}}D_k^\mu(t), \hspace{32pt} k \geq 1  \label{eq:KLocal}\\
	&= \frac{2^{|C|}(-1)^{k}}{2^{|C|} - 1}  \sum_{\substack{A \subseteq S \\ |A| \leq k}} (-2)^{|A|} {N - |A| \choose N - k} \Tr[\rho^{AC}(t)^2] \nonumber
	\end{align}
	We prove the second equation in Appendix \ref{app:KLocal}. The above quantity allows one to track how operators initially located within $C$ increase in complexity (in the sense of $k$-locality) with time.
	Later, we will use $D_k^C(t)$ as a means to quantify this aspect of operator growth on a quantum device.\\
	
	In the following section, we demonstrate that the quantities described above, which depend only on integer moments of the operator state $\rho_{\rm op}(t)$, can be directly measured in experiment without using full tomography. Moreover, this can be done without ever explicitly constructing the doubled state, which would require simultaneous access to identical copies of the system.
	
	\section{Shadow tomographic measurement of operator-space entanglement \label{sec:Protocol}}
	
	The method we use to measure operator-space R{\'e}nyi entropies is based on classical shadow tomography \cite{Huang2020}. There, one performs projective measurements in different randomly selected bases on a target state $\rho$, each of which gives a particular snapshot of $\rho$. The ensemble of snapshots (known as the `shadow' of $\rho$) has an efficient classical representation, which allows one to calculate estimators of expectation values $\Tr[O \rho]$ and non-linear moments $\Tr[A \rho^{\otimes m}]$ using classical post-processing on the shadow data.
	
	Here, we propose to build up a shadow of the doubled state $\rho_{\rm op}(t)$ by preparing random states, evolving them under $\mathcal{N}_t$, and performing measurements in independently chosen random bases. For our purposes, the random states and bases will be related to the computational basis by single-qubit rotations, since these can be implemented accurately on current devices; however generalizations to global rotations are also possible \cite{Huang2020,Hu2021}.
	
	\begin{figure}
		\includegraphics[width=246pt]{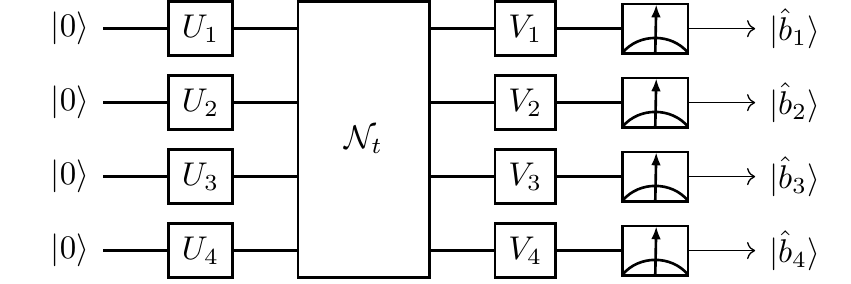}
		\caption{Illustration of experimental protocol to measure operator-space entanglement of a quantum channel $\mathcal{N}_t$ in a system with $N = 4$ qubits. The single qubit unitaries $U_j$, $V_j$ are drawn independently at random from the discrete gate sets described in the main text. Once the measurement outcomes $\hat{b}_j$ are known, one can construct a snapshot of the doubled state $\rho_{\rm op}(t)$ using Eq.~\eqref{eq:RhoEst}, and then repeat $M$ times with different unitaries.}
		\label{fig:Protocol}
	\end{figure}
	
	The specific protocol is illustrated in Fig.~\ref{fig:Protocol}. Output rotations $V_j$ applied immediately prior to measurement are sampled independently from a uniform distribution over the discrete set of gates $\{\mathbb{I}, H_X, H_Y\}$, where $H_{X, Y}$ are $X$- and $Y$-Hadamard gates. This effectively implements one of the 3 possible Pauli measurements for each qubit. The gates $U_j$ applied prior to time evolution are chosen such that the distribution of initial input states $U_j\ket{0}$ is uniform over the 6 states $\{\ket{\pm_{\sigma}} : \sigma = X, Y, Z\}$, where $\ket{+_\sigma}$ ($\ket{-_\sigma}$) is the eigenstate of the Pauli operator $\sigma$ with eigenvalue $+1$ ($-1$). A total of $M$ runs are performed, and for now we assume that a new set of independent gates are generated for each run.
	
	The data associated with a particular run are the gates $U_j$, $V_j$, along with the measurement outcomes $\hat{b}_j \in \{0, 1\}$. These can be used to construct a snapshot of $\rho_{\rm op}(t)$ (we use a hat to distinguish this estimator from the true operator state)
	\begin{align}
	\hat{\rho}_{\rm op}(t) &= \bigotimes_{j=1}^N \Big(3U_j^T \ket{0}\bra{0} U_j^* - \mathbb{I}\Big)_{\rm in} \nonumber\\ &\otimes \bigotimes_{j=1}^N  \Big(3V_j \ket{\hat{b}_j}\bra{\hat{b}_j} V_j^\dagger - \mathbb{I}\Big)_{\rm out}.
	\label{eq:RhoEst}
	\end{align}
	Using the arguments of Ref.~\cite{Huang2020}, along with the definition of $\rho_{\rm op}(t)$ and the property of the maximally entangled state $(O \otimes \mathbb{I})\ket{\Phi} = (\mathbb{I} \otimes O^T)\ket{\Phi}$, one can show that the above is an unbiased estimator of $\rho_{\rm op}(t)$, i.e.~$\mathbb{E}[\hat{\rho}_{\rm op}(t)] = \rho_{\rm op}(t)$, where the expectation value is over both random unitaries $U_j, V_j$ and measurement outcomes $\hat{b}_j$. (See Appendix \ref{app:RhoEst}. Eq.~\eqref{eq:RhoEst} is consistent with other similar proposals that have appeared recently \cite{Levy2021,Kunjummen2021}.) A different snapshot is obtained from each of the $M$ runs, and we write the snapshot obtained from the $r$th run as $\hat{\rho}_{\rm op}^{(r)}(t)$.
	
	For each $\hat{\rho}_{\rm op}^{(r)}(t)$, an independent unbiased estimator of a given correlation function $\Tr[O_{\rm in} O_{\rm out}(t)]$ can be constructed by computing $\Tr[(O_{\rm in} \otimes O_{\rm out}) \hat{\rho}_{\rm op}^{(r)}(t)]$ on a classical computer. For sufficiently large $M$, the average over all estimators gives an accurate prediction of the correlation function. Estimators for non-linear functionals, such as the moments $p_{m, :AC} = \Tr[\rho^{AC}(t)^m]$ appearing in the R{\'e}nyi entropies, can be constructed using so-called $U$-statistics \cite{Ferguson2003}, as one does in conventional shadow tomography. For instance, to estimate $p_{2,:AC}$, one can average $\Tr[\hat{\rho}^{AC, (r_1)}(t) \hat{\rho}^{AC, (r_2)}(t)]$ over all $M(M-1)$ ordered pairs of independent snapshots $r_1 \neq r_2$, where $\rho^{AC, (r)}(t) \coloneqq \Tr_{B \cup D} \hat{\rho}_{\rm op}^{(r)}(t)$. The snapshot \eqref{eq:RhoEst} can also be partially transposed beforehand to obtain $p_{m, A:C}$. Conveniently, the same set of shadow data can be used to obtain multiple quantities simply by post-processing in different ways.
	
	The size of the statistical errors that arise from this process will depend on the particular quantity being estimated, the channel $\mathcal{N}_t$ in question, and the sample count $M$. Worst-case upper bounds on the number of samples $M_\epsilon$ required to achieve an error $\epsilon$ in state shadow tomography have been derived in Refs.~\cite{Huang2020,Elben2020}, and these can be carried over to the present setting, at least for single-qubit rotations. For the moments $p_{m, :AC} = \Tr[\rho^{AC}(t)^m]$ (with or without partial transposition) in the small-$\epsilon$ limit, one has $M_\epsilon \leq \mathcal{O}(2^{|AC|}/\epsilon^2)$. In the Supplemental Material, we argue that when $\rho^{AC}(t)$ is highly mixed (which is common for operator-space states), a potentially tighter upper bound of $\mathcal{O}(2^{|AC|\log_2 3 - S^{(\infty)}({AC})}/\epsilon^2)$ applies, where $S^{(\infty)}(AC) = -\log \max {\rm eig}\, \rho^{AC}(t)$ is the max-entropy. While this is exponential in the number of qubits in $AC$, the scaling is highly favourable over the $\mathcal{O}(2^{|AC|} \text{rank}(\rho_{AC})^2/\epsilon^2\log(1/\epsilon))$ number of runs required for full tomography using the same resources (i.e.~only single-qubit rotations) \cite{Haah2017}. In general, while these bounds are expected to have the correct scaling behaviour, the prefactors involved are typically not tight \cite{Huang2020}.

	\section{Simulating and detecting quantum chaos \label{sec:IBM}}
	
	We now present results of simulations of quantum chaotic dynamics performed on a cloud-based IBM superconducting quantum processor, using the method described above to access operator-space measures of scrambling. The system in question, \texttt{ibm\_lagos} \cite{IBMQ}, has 7 qubits, arranged as illustrated in Fig.~\ref{fig:RenyiMI}(a). In the main text, we present results where 5 contiguous qubits are used to simulate a 1D chaotic system using entangling gates arranged in a brickwork pattern [Fig.~\ref{fig:RenyiMI}(b)]. Appendix \ref{app:N7} contains details of similar results that involve all 7 qubits in the device, for which an alternative spacetime pattern of gates is needed.
	
	\subsection{Setup}
	
\begin{figure}
		\centering
		\includegraphics[width=246pt]{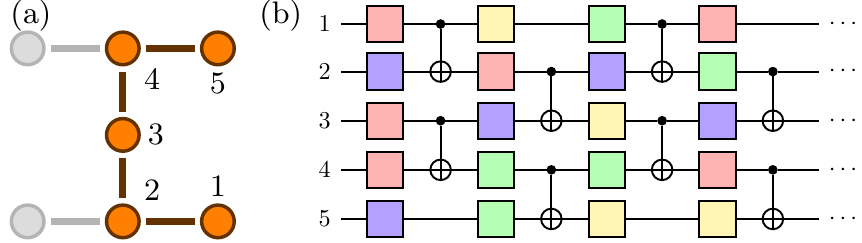}
		\caption{(a) Qubit layout and connectivity of \texttt{ibm\_lagos}. Dark purple circles represent the 5 qubits used for the experiments detailed in the main text. (b) Circuit design for the chaotic unitary $\mathcal{N}_t$, with $t = 4$ timesteps shown. Each single qubit gate (coloured boxes) is independently sampled from the four gates $W_{1, \ldots, 4}$, see main text.}
		\label{fig:RenyiMI}
	\end{figure}

	The brickwork circuit is made up of entangling two-qubit gates, which we choose to be CNOTs, combined with single-site unitaries. Each single-site gate is independently sampled from a uniform distribution over a discrete set of 4 gates $\{W_{c}: c = 1, \ldots, 4\}$. In terms of the native gates of the quantum device ($\sqrt{X}$, $X$, and $R_\theta = e^{-\iu \theta Z/2}$), these are $W_1 = R_{\pi/4} \sqrt{X} R_{\pi/4}^\dagger,\, W_2 = R_{\pi/4} X R_{\pi/4}^\dagger,\, W_3 =  \sqrt{X} R_{\pi/4} \sqrt{X} ,\, W_4 = \sqrt{X} R_{\pi/4}^\dagger \sqrt{X}$. In a given timestep $t = 1, 2, \ldots$, CNOTs are applied to pairs of qubits $(2j-1, 2j)$ for odd $t$ (first index is control, second is target), and to $(2j, 2j+1)$ for even $t$. All qubits are then subjected to single-site unitaries $W_{c_{j,t}}$. This circuit is illustrated in Fig.~\ref{fig:RenyiMI}(b). The indices $c_{j,t}$ have been sampled once for each $j, t$, and this configuration is used in all the data presented in this paper, i.e.~we do not average over different single-qubit unitaries. This defines a time-dependent evolution channel $\mathcal{N}_{t = 1, 2, \ldots}$ that exhibits chaos.
	
	In practice, for the quantum processor we use, running the same circuit many times is much faster than running many randomly generated circuits once each. For this reason, we alter the shadow protocol slightly: A random computational basis state $\ket{\psi} = \bigotimes_{j=1}^N \ket{\hat{a}_j}$ is used in place of the initial $\ket{0^{\otimes N}}$ (this can be done with a fixed circuit by preparing $\ket{0^{\otimes N}}$ and applying Hadamard gates to each qubit, followed by projective measurements of all qubits). The full circuit is sampled $M_{\rm S}$ times for a fixed choice of $U_j$, $V_j$, generating different $\hat{a}_j$, $\hat{b}_j$ each time. The whole procedure is repeated for $M_{\rm U}$ different independently chosen bases. While the sampling errors in the final outcomes of observables are sub-optimal for a fixed total number of runs $M_{\rm S}M_{\rm U}$ compared to the usual shadows protocol \cite{Huang2020}, we are able to reach a much higher total run count this way, thus achieving higher accuracy. We discuss the necessary alterations to the post-processing methods and the influence on the scaling of errors in the Supplemental Material \cite{SM}.
	
	Other than the channel $\mathcal{N}_t$ itself, the full circuit involves single-qubit unitaries and measurements. To compensate for the imperfect measurement process, we ran periodic calibration jobs, the data from which was used to apply measurement error mitigation techniques as described in e.g.~Ref.~\cite{Qiskit}. In principle, one could also employ a version of shadow tomography that counteracts the effects of errors in the unitaries $U_j$, $V_j$ \cite{Chen2021}; however, the single-qubit gate errors in \texttt{ibm\_lagos} are on the order of $10^{-4}$, so we assume that these unitaries are implemented perfectly.

	The full shadow tomography protocol was executed on \texttt{ibm\_lagos} with $M_S = 8192$, $M_U = 900$, for $t$ varying from 0 to 15. The values obtained from this dataset are affected by both imperfections in $\mathcal{N}_t$ realised in the quantum device  (`noise') and the sampling error (i.e.~the statistical fluctuations arising from the stochastic nature of shadow tomography). To help distinguish these two sources of error, we have also generated another set of shadow data by running noise-free numerical simulations of the full circuits [Fig.~\ref{fig:Protocol}] where all gates are perfectly accurate, and the measurement outcomes are sampled stochastically. This dataset generates values that are affected by sampling error only. The same two sets of shadow data (which we label `simulation' and `\texttt{ibm\_lagos}') were used to calculate all the different physical quantities described in the following. We also compute the exact value of each quantity for noiseless $\mathcal{N}_t$, against which the shadow tomographic estimates will be compared.
	 Throughout, we fix $A$ and $C$ to be individual qubits, $A = \{1\},\, C = \{j_C\}$, where $j_C = 1, \ldots, N$.
	
	\subsection{Results}
	
	\begin{figure}
		\centering
		\includegraphics*[width=246pt]{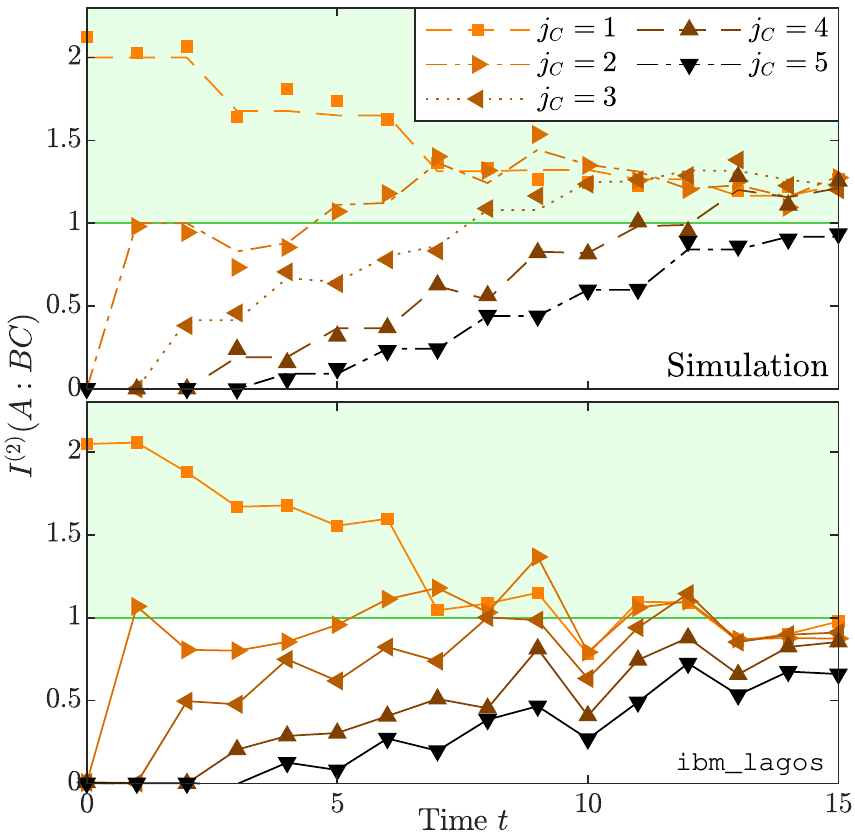}
		\caption{R{\'e}nyi mutual information [Eq.~\eqref{eq:RenyiMutual}], with $A = \{1\},\, C = \{j_C\}$. Top panel: Dashed lines indicate the exact value without noise or sampling error, points are estimations obtained using shadow post-processing methods on data from numerical simulations of the full circuit (Fig.~\ref{fig:Protocol}) without noise. The deviations between these two values can be used to estimate the typical size of the sampling errors that arise from the shadow tomography protocol. Bottom panel: results obtained from \texttt{ibm\_lagos}; solid lines are to guide the eye. The region above the threshold $I^{(2)}(A:BC) > 1$ is shaded green (see Section \ref{sec:RenyiScrabling}).}
		\label{fig:Mutinf5}
	\end{figure}
	
	Firstly, the R{\'e}nyi mutual information $I^{(2)}(A:BC)$ is plotted in Fig.~\ref{fig:Mutinf5}. At early times, the mutual information is large only for $j_C = 1$, reflecting the fact that the input $A$ can only be reconstructed if one has access to the same qubit at the final time. At late times, the data from noiseless simulations saturate to comparable values for all choices of $j_C$, close to the value $I^* = 1.1945\ldots$ that would be expected if $\mathcal{N}_t$ were a global Haar random unitary (see Appendix \ref{app:RenyiMI}), thus confirming that information has scrambled. (For $j_C = 5$, this value is reached at a time just beyond the maximum $t$ simulated on the quantum device.) The approach to this saturation value follows a light-cone structure: qubits that are further away from $A$ take a longer time to reach saturation. The results from the quantum processor agree well with simulations at early times. At later times we see an increasingly marked reduction of $I^{(2)}(A:BC)$ for all $j_C$. This is a consequence of the cumulative effects of noise in the execution of the time evolution $\mathcal{N}_t$, which reduces the fidelity of information transmission. For $j_C \leq 3$, we find values of $I^{(2)}(A:BC)$ above the threshold value of $|A| = 1$, which confirms that the quantum communication capacity of $\mathcal{N}^{A \rightarrow BC}_t$ is non-zero (see previous section). Even though the threshold is not exceeded for all qubits due to noise, the increase of $I^{(2)}(A:BC)$ confirms that information does indeed propagate to all qubits to some extent.
	
	\begin{figure}
		\includegraphics[width=246pt]{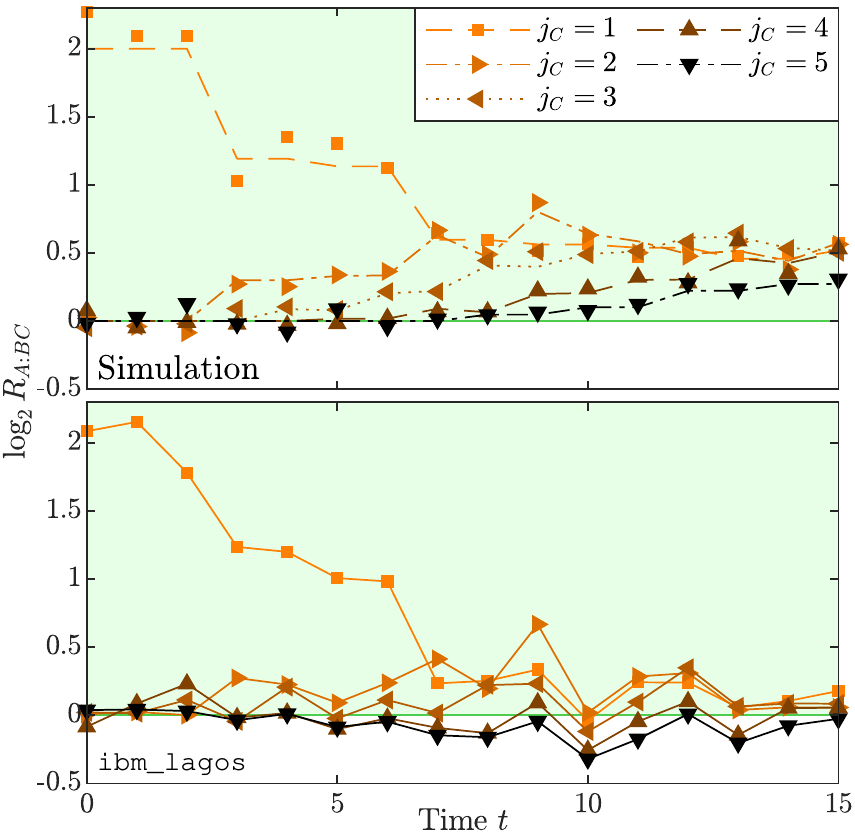}
		\caption{Logarithm of the ratio $R_{A:BC} = p_{2,A:BC}^2/p_{3,A:BC}$, where $A = \{1\}$, $C = \{j_C\}$. Data presented as in Fig.~\ref{fig:Mutinf5}. The region above the threshold $\log R_{A:BC} > 0$ is shaded green (see Section \ref{sec:RenyiScrabling}).}
		\label{fig:Negativity}
	\end{figure}
	
	The ratio of negativities $R_{A:BC}$ is plotted in Fig.~\ref{fig:Negativity}. These show a similar pattern to the mutual information: The early-time values of $R_{A:BC}$ are large only for $j_C = 1$, and as time evolves the ratio tends towards saturation values that are comparable for all values of $j_C$, following a light-cone structure. From numerical simulations, we see that the threshold $R_{A:BC} > 1$ is achieved at earlier times than for the R{\'e}nyi mutual information, suggesting that this criterion is more sensitive than the mutual information to the particular form of operator-space entanglement generated by the dynamics. On the other hand, the data from \texttt{ibm\_lagos} shows a more significant suppression of the signal, suggesting that the quantity in question may be more sensitive to noise.
	
	\begin{figure}
		\includegraphics[width=246pt]{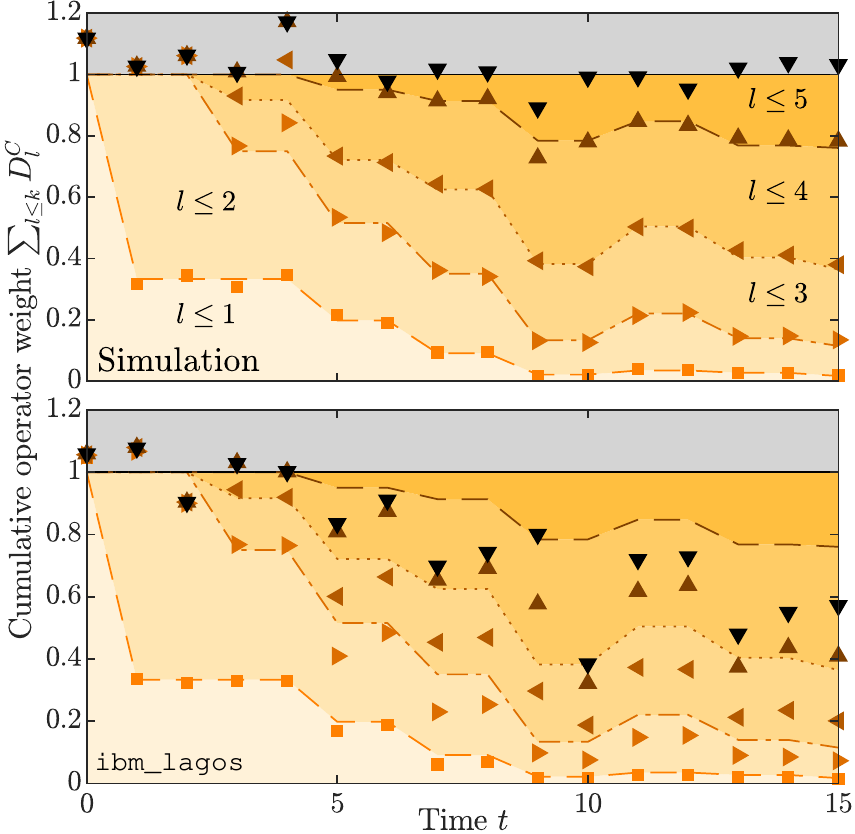}
		\caption{Evolution of the $k$-locality of time-evolved  operators, as quantified by $D_k^C$ [Eq.~\eqref{eq:KLocal}]. Specifically, we plot the cumulative weight $\sum_{l\leq k} D_l^C$ which measures the total weight of the time-evolved operator acting non-trivially on at most $k$ qubits, averaged over all non-trivial initial operators with support on $C$. We fix $C = \{3\}$, the central qubit in Fig.~\ref{fig:RenyiMI}(a). The shaded areas and dashed lines indicate the exact values without sampling error or noise. Markers indicate shadow tomographic estimates calculated from the datasets obtained from noise-free simulations (top) and from the quantum device (bottom). The former are affected by sampling error only, while the latter are affected by both sampling error and noise.}
		\label{fig:Opweights}
	\end{figure}
	
	Finally, in Fig.~\ref{fig:Opweights}, we plot the cumulative sums $\sum_{l = 0}^k D_l^C$ [Eq.~\eqref{eq:KLocal}], which measures the proportion of the time-evolved operators $\pstring^\mu(t)$ that act non-trivially on at most $k$ qubits, averaged over all non-identity initial operators $\pstring^\mu$ with support in $C$. Here we fix $C = \{3\}$, the central qubit in the chain. Note that for unitary time evolution, the total operator weight $\sum_\nu |c^{\mu \nu}(t)|^2$ is conserved, which implies that $\sum_{l=0}^N D_l^C = 1$.
	
	At early times, the operators have only evolved a small amount away from their single-qubit initial values, and so the operator weight is dominated by the low-$k$ sectors. As time evolves, an increasing amount of weight moves onto operators with more extended support. Eventually, once the system has fully scrambled, the evolved operators have weight roughly evenly distributed over the whole space of operators (excluding identity). The weights $D_k^C(t)$ are then well approximated by $(4^N - 1)^{-1} 3^k {N \choose k}$, which is the value that would be obtained from a uniform distribution over all $4^N - 1$ non-trivial operators. At these late times, the values obtained from \texttt{ibm\_lagos} are again lower than the exact values due to noisy non-unitary processes. Indeed, given that the dynamics of the quantum device is not perfectly unitary, the total operator weight $\sum_\nu |c^{\mu \nu}(t)|^2$ is expected to decrease with time, which is reflected in the data for $k = 5$.

	\section{Discussion and Outlook \label{sec:Discussion}}
	
	Using a combination of randomized state preparation and measurement, combined with the postprocessing techniques introduced in Ref.~\cite{Huang2020}, we have evaluated various operator-space entanglement measures in a programmable quantum simulator. We constructed quantities that probe the fidelity of the Hayden-Preskill teleportation protocol \cite{Hayden2007}, allowing us to unambiguously confirm that the system exhibits scrambling. Additionally, we used the same techniques to characterise operator growth, which can also be used to diagnose quantum chaos \cite{Nahum2018, vonKeyserlingk2018}.
	
	A related approach to diagnosing scrambling in experiments is to measure the decay of OTOCs \cite{Li2017, Garttner2017, Wei2018, Joshi2020, Mi2021, Garcia2021}. However, present day quantum simulators are inevitably noisy, and dissipative effects can mimic this decay \cite{Yoshida2019, Zhang2019}, as can mismatch between forward and backward time evolution. Thus, OTOC decay is at present not a truly verifiable diagnostic of scrambling to the same extent as many-body teleportation.
	
	Compared to previous proposals to measure operator-space entanglement and teleportation fidelities \cite{Landsman2019,Sun2020}, our method has the advantage that no additional ancilla qubits are needed. Not only does this reduce the hardware requirements in terms of system size, it also removes the need to control the dynamics of ancillas, which would otherwise need to be kept coherent, and possibly time-evolved in parallel \cite{Yoshida2019}. Moreover, other than the time evolution $\mathcal{N}_t$ itself, the only additional gates required are single-qubit rotations, making the protocol particularly straightforward to implement on a wide variety of programmable quantum simulators. This simplicity is possible because our protocol does not require us to explicitly perform the decoding procedure for the many-body teleportation problem; rather, we can infer the existence of correlations between $A$ and $BC$ from statistical correlations between different measurement, which in turn informs us that teleportation is in principle possible.
	
	In developing the protocol used here, we have focussed on keeping experimental requirements to a minimum. However, other approaches that demand higher levels of experimental control may offer different advantages. In particular, one consequence of using randomized state preparation and measurement is the exponential scaling of the required number of repetitions $M$ with the size of the region on which the R{\'e}nyi entropy is evaluated --- indeed, this sampling complexity is provably optimal with the given resources \cite{Huang2020}.
	This is not an issue if one is interested in small regions within a large system, which is the situation for many studies of quantum thermalization, but may be problematic if one needs to consider large $AC$. Indeed, the ideal probes of many-body teleportation require access to an extensive number of inputs $AB$. Note, however, that one could consider correlations between $A$ and $B'C$, where $B'$ is a fixed size rather than the full complement of $A$, which will be good measures of early-time chaos; see also the modified OTOCs in Ref.~\cite{Vermersch2019}.
	
	One immediate generalization is to replace the random local unitaries $U_j$, $V_j$ with global Clifford gates \cite{Huang2020}. As argued in Ref.~\cite{Elben2018}, the scaling of the required number of runs will be better, albeit still exponential. The larger number of gates required will make such a protocol more susceptible to decoherence, and so noise-robust techniques would be required \cite{Chen2021}.
	
	If the evolution in question $\mathcal{N}_t$ is known in advance, then further improvements to the scaling of $M$ may be obtained using ancillary qubits. Roughly speaking, in these approaches the non-local correlations established during time evolution are distilled into smaller regions using some decoding procedure that requires knowledge of $\mathcal{N}_t$; these correlations can then be verified in a sample-efficient way. For instance, fast decoders for the Hayden-Preskill problem have been developed that use a doubled system \cite{Yoshida2017}. Note that as the system size increases, so too will the complexity of these decoders, requiring increasingly high levels of coherence and gate fidelity. Thus, in current NISQ devices, there is a natural tradeoff between sample complexity and the necessary level of control over the system.
	
	The quantities that one can directly access without using full tomography of $\mathcal{N}_t$ or an ansatz for $\rho^{AC}(t)$ \cite{Kokail2021} are limited to integer moments of the (doubled) density matrix $\rho_{\rm op}(t)$. While the R{\'e}nyi entropies $S^{(m)}(AC)$ and partially transposed moments $p_{m, A:C}$ have less information-theoretic significance than, e.g.~the von Neumann entropy, their experimental relevance makes it important to better understand their behaviour in chaotic systems, which we leave to future work.
	
	Recently, a protocol to measure the spectral form factor --- a quantity that can be used to diagnose chaos in time-periodic systems \cite{Haake2010} --- has been proposed, which also uses randomized state preparation and measurement \cite{Joshi2022}. There, the initial and final unitaries appearing in Fig.~\ref{fig:Protocol} are related via $U_j = V_j^\dagger$. It would be interesting to consider other ways of introducing correlations between different random unitaries in such protocols, which could give access to different properties of the time-evolution channel.
	
	Operator-space entanglement also plays an important role in contexts beyond quantum chaos. For instance, the mutual information between initial and final states can be used as a probe of entanglement phase transitions in monitored quantum circuits \cite{Li2018,Li2019,Bao2020,Gullans2020}. Analogous quantities can also be used to detect quantized chiral information propagation at the edge of anomalous Floquet topological phases \cite{Rudner2013,Po2016, Duschatko2018, Gong2021}. The protocol we employ here could therefore be used as a means to verify experimental realisations of these phenomena.\\
	
	\textit{Note added.---}During completion of this work, Refs.~\cite{Levy2021,Kunjummen2021} appeared, where similar proposals to generalize shadow tomography to channels were given.
	
	\begin{acknowledgments}
		We acknowledge support from EPSRC Grant EP/S020527/1. We acknowledge the use of IBM Quantum services for this work. SJG is supported by the Gordon and Betty Moore Foundation. JJ is supported by Oxford-ShanghaiTech collaboration agreement. The views expressed are those of the authors, and do not reflect the official policy or position of IBM or the IBM Quantum team. Statement of compliance with EPSRC policy framework on research data: Data obtained from numerical simulations and experiments on \texttt{ibm\_lagos} will be made publicly accessible via Zenodo upon publication.
	\end{acknowledgments}

	\appendix
	\section{Properties of the R{\'e}nyi mutual information \eqref{eq:RenyiMutual} \label{app:RenyiMI}}
	
	In this section, we prove the claims made in the main text regarding properties of R{\'e}nyi mutual information [Eq.~\eqref{eq:RenyiMutual}] when the underlying state is an operator state $\rho_{\rm op}(t)$, including our claim that the quantum capacity of a channel must be non-zero when the corresponding operator-space R{\'e}nyi mutual information exceeds its maximum classical value. We will refer explicitly to the quantity $I^{(m)}(A:C) \coloneqq S^{(m)}(A) + S^{(m)}(C) - S^{(m)}(AC)$ where $A$ is a subset of inputs and $C$ is a subset of outputs [as in Fig.~\ref{fig:HaydenPreskill}(b)]; however our claims continue to hold if $A$ and $C$ are replaced by subsets that contain combinations of inputs and outputs, provided that the reduced density matrix on at least one of the subsets is maximally mixed. For example, in the main text we consider $I^{(m)}(A:BC)$, which falls under this category since the reduced density matrix $\rho^A =  \mathbb{I}_A/2^{|A|}$ is maximally mixed. For the purposes of this appendix, we leave all $t$-dependence implicit. We denote the Hilbert space dimensions of $A$, $C$ as $d_{A}$, $d_C$, respectively.\\
	
	While the definition of the R{\'e}nyi mutual information that we use here [Eq.~\eqref{eq:RenyiMutual}] generalizes the von Neumann mutual information in a natural way, it is not always a good measure of the correlations present in a given state. For example, in certain cases it can even be negative \cite{Berta2015,Scalet2021}. (Because of this, other related quantities have been proposed that are sometimes referred to as R{\'e}nyi mutual information \cite{Wilde2014}; here we will use this term exclusively for the quantity \eqref{eq:RenyiMutual}.) However, when the reduced density matrix for either $A$ or $C$ is maximally mixed -- as occurs in the cases under consideration -- it was noted that $I^{(m)}(A:C)$ is non-negative \cite{Lensky2019}. We argue that this can be made stronger:
	\begin{theorem*}
		For any density operator $\rho^{AC}$ satisfying $\rho^A \coloneqq \Tr_C \rho^{AC} = \mathbb{I}_A/d_A$ or $\rho^C = \mathbb{I}_C/d_C$, the R{\'e}nyi mutual information satisfies
		\begin{align}
		I^{(m)}(A:C) &\geq 0 & \forall m = 2, 3, \ldots
		\label{eq:MINonNegative}
		\end{align}
		with equality if and only if the density operator factorizes as $\rho^{AC} = \rho^A \otimes \rho^C$. 
	\end{theorem*}
	This theorem establishes $I^{(m)}(A:C)$ as a sensible measure of how much $\rho^{AC}$ fails to factorize, and hence the degree to which $A$ and $C$ are correlated. We only explicitly consider integer $m \geq 2$ here, since these are the quantities that can be measured experimentally.
	
	We assume that $\rho^A$ is maximally mixed; the alternative case where $\rho^C$ is maximally mixed then follows from the symmetry of $I^{(m)}(A:C)$. Our proof relies on the following observation
	\begin{align}
	&\int_{\rm Haar} \dif U_1 \cdots \dif U_m \Tr\Big[(U_1 \otimes \mathbb{I}_C) \rho_{AC} (U_1 \otimes \mathbb{I}_C)^\dagger \cdots \nonumber\\ & \cdots(U_m \otimes \mathbb{I}_C) \rho_{AC} (U_m \otimes \mathbb{I}_C)^\dagger\Big] = \Tr\big[(\mathbb{I}_A/d_A)^m \otimes (\rho_C)^m\big]
	\end{align}
	where the integration variables are unitary matrices $\{U_i \in \mathrm{U}(d_A)\}$ acting on $A$, and the integrals are taken over the Haar measure. The above is a consequence of the standard identity $\int_{\rm Haar} \dif U\, U O U^\dagger = (\Tr[O]/d)\, \mathbb{I}_d$ for $d \times d$ matrices $O$ \cite{Collins2006}. We seek to prove $\Tr[(\rho^{AC})^m] \geq \Tr[(\rho^{A})^m \otimes (\rho^C)^m]$, which will in turn imply \eqref{eq:MINonNegative}. Since the integration measure over each $U_i$ is normalized $\int_{\rm Haar} \dif U_i = 1$, and $\rho^A = \mathbb{I}_A/d_A$, we have
	\begin{align}
	&\Tr[(\rho^{AC})^m] - \Tr[(\rho^{A})^m \otimes (\rho^C)^m]  = \int_{\rm Haar} \dif U_1 \cdots \dif U_m \nonumber\\ & \bigg(\Tr[(\rho^{AC})^m] -
	\Tr\Big[U_1 \rho^{AC} U_1^\dagger \cdots U_m  \rho^{AC} U_m^\dagger\Big]\bigg)
	\label{eq:TraceIntegral}
	\end{align}
	where we leave the factors of $\mathbb{I}_C$ implicit. The integrand of the right hand side is non-negative by the following lemma
	\begin{lemma*}
		If $\rho$ is a complex Hermitian positive semi-definite matrix and $\{U_i\}$ are unitary matrices of the same size, then
		\begin{align}
		\big|\!\Tr[U_1 \rho U_1^\dagger \cdots U_m \rho U_m^\dagger]\,\big| \leq \Tr[\rho^m]
		\label{eq:TraceIneq}
		\end{align}
		with equality if and only if $U_1 \rho U_1^\dagger = \cdots = U_m \rho U_m^\dagger$.
	\end{lemma*}
	\textit{Proof.---} We first note that $|\!\Tr[A]| \leq \Tr[|A|]$ for all square matrices $A$, where $|A| \coloneqq (A^\dagger A)^{1/2}$, with equality if and only if $A$ is Hermitian positive semidefinite. Setting $A = A_1 \cdots A_m$ where $A_j = U_j \rho U_j^\dagger$, we then use a generalization of H{\"o}lder's inequality proved in Ref.~\cite{Manjegani2007}:
	\begin{align}
	\Tr[|A_1 \cdots A_m|] \leq \prod_{a=1}^m \Big(\!\Tr[(A_a)^{p_a}]\Big)^{1/p_a}
	\label{eq:GeneralizedHolder}
	\end{align}
	for any positive real numbers $p_a$ satisfying $\sum_a (1/p_a) = 1$, with equality if and only if $A_1 = \cdots = A_m$. Eq.~\eqref{eq:TraceIneq} then follows by setting $p_a = m$ for $a = 1,\ldots, m$, so that $\Tr[(A_a)^{p_a}] = \Tr[\rho^m]$. \hfill $\blacksquare$\\
	
	This completes our proof that $I^{(m)}(A:C)$ is non-negative for the states under consideration. The fact that $I^{(m)}(A:C)$ vanishes for factorizable $\rho^{AC}$ follows immediately from its definition. Conversely, if $I^{(m)}(A:C) = 0$, then the integrand in \eqref{eq:TraceIntegral} must vanish everywhere, which implies that $(U \otimes \mathbb{I}_C) \rho^{AC} (U \otimes \mathbb{I}_C)^\dagger = \rho^{AC}$ for all $U \in \mathrm{U}(d_A)$. This can only be true if $\rho^{AC} \propto \mathbb{I}_A \otimes \rho^C$, which completes our proof.\\
	
	Having established the above theorem, we now provide the proof of the claims we made in Section \ref{sec:RenyiScrabling} regarding the threshold values for $I^{(m)}(A:BC)$ and $R_{A:BC}$. In its most general form, we have
	\begin{claim*}
		The quantum capacity of a channel $\mathcal{N}^{A \rightarrow B}$ is non-zero if the operator-space R{\'e}nyi mutual information satisfies $I^{(m)}(A:B) > |A|$. If the input Hilbert space dimension $|\mathcal{H}_A| = 2$, then the same conclusion can be made whenever the ratio of partially transposed moments $R_{A:B} \coloneqq p_{2, A:B}^2/p_{3,A:B}$ exceeds unity.
	\end{claim*}
	The statements made in the main text then follow from applying the above to $\mathcal{N}_t^{A\rightarrow B'C}$.\\
	
	\textit{Proof.---}Firstly, we consider the case where $I^{(m)}(A:B) > |A|$. Here we will rely somewhat on the notion of majorization; see, e.g.~Ref.~\cite{Marshall1979} for a full introduction. A $n_X \times n_X$ Hermitian matrix $X$ majorizes a $n_Y \times n_Y$ Hermitian matrix $Y$ if their traces are equal and the sum of the $k$th largest eigenvalues of $X$ is greater than or equal to the sum of the $k$th largest eigenvalues of $Y$ for $k = 1, 2, \ldots, \min(n_X, n_Y)$. This relation is denoted denoted $X \succeq Y$. A function $f$ from matrices to real numbers is called Schur convex iff $X \succeq Y \Rightarrow f(X) \geq f(Y)$.
	
	Since $A$ is maximally mixed, our starting point $I^{(m)}(A:B) > |A|$ is equivalent to $\Tr[(\rho^{AB})^m] > \Tr[(\rho^{B})^m]$, where $\rho^{AB}$ is the operator state for the channel $\mathcal{N}^{A \rightarrow B}$ (see Eq.~\ref{eq:RhoOp}), and $\rho^B = \Tr_A \rho^{AB}$. It is straightforward to show that the map $\rho \mapsto \Tr[\rho^m]$ is Schur-convex, which implies that $\rho^{AB} \npreceq \rho^{B}$. In Ref.~\cite{Nielsen2001}, it was shown that separable states satisfy $\rho^{AB} \preceq \rho^B$, and so the operator-state must be bipartite entangled whenever $I^{(m)}(A:B) > |A|$. Moreover, in Ref.~\cite{Hiroshima2003} a stronger result was proved: violation of the separability criterion $\rho^{AB} \preceq \rho^B$ implies violation of the so-called reduction criterion \cite{Horodecki1999}. States which violate the reduction criterion must possess \textit{distillable} entanglement, meaning that many copies of the state can be converted into a smaller number of pure EPR pairs using local operations and classical communication \cite{Horodecki1998}.
	
	The above implies that if the operator-state $\rho^{AB}$ satisfies $I^{(m)}(A:B) > |A|$, then pure EPR pairs can be distilled from many copies of $\rho^{AB}$ (each of which can be prepared from a single use of the channel $\mathcal{N}^{A\rightarrow B}$) using the protocol described in Ref.~\cite{Horodecki1999}, which requires a one-way classical communication channel from sender $A$ to receiver $B$. The ability to generate EPR pairs from multiple uses of a channel assisted by one-way classical communication is equivalent to being able to reliably transmit the same number of qubits from $A$ to $B$ using the same resources \cite{Bennett1996}. Since the quantum channel capacity assisted by one-way classical communication is equal to the unassisted capacity \cite{Bennett1996, Barnum2000}, we conclude that the quantum capacity of any channel $\mathcal{N}^{A \rightarrow B}$ must be non-zero whenever the operator-state $\rho^{AB}$ satisfies $I^{(m)}(A:B) > |A|$.\\
	
	For the ratio of partially transposed moments $R_{A:B}$ [Eq.~\eqref{eq:RRatio}], our argument follows a similar line. In Ref.~\cite{Elben2020}, it was shown that if a bipartite state $\rho^{AB}$ satisfies $R_{A:B} > 1$, then the Peres criterion \cite{Peres1996} must be violated, which is a sufficient but not necessary condition for the existence of bipartite entanglement in $\rho^{AB}$. Given that the Hilbert space dimension $|\mathcal{H}_A| = 2$, violation of the Peres criterion implies that the entanglement in $\rho^{AB}$ is distillable \cite{Dur2000}. Again using the equivalence between generation of pure EPR pairs and transmission of quantum states, we conclude that the quantum capacity of $\mathcal{N}^{A \rightarrow B}$ must be non-zero.\\
	
	Finally, it is helpful to evaluate $I^{(m)}(A:C)$ for the case where the time evolution is a global Haar-random unitary, which is maximally chaotic. A simple estimate for the average $\langle I^{(m)}(A:C) \rangle_{U_t}$ (angled brackets denote the expectation value over all unitary evolutions $U_t$ with respect to the Haar measure) can be obtained by approximating $\langle \log \Tr[\rho^{AC}(t)^m] \rangle_{U_t} \approx \log \langle \Tr[\rho^{AC}(t)^m] \rangle_{U_t}$, the right hand side of which can be evaluated using standard expressions for integrals over the Haar measure \cite{Collins2006}. This assumes that fluctuations of $\Tr[\rho^{AC}(t)^m]$ between different Haar-random unitaries are small. For the simplest case of $m = 2$, for a system of $N$ $q$-level systems ($q = 2$ for our case of qubits), we find
	\begin{align}
		\langle \Tr[\rho^{AC}(t)^2] \rangle_{U_t} &= \frac{1}{q^N(q^{2N}-1)}\bigg[ q^{N}(q^{|BD|}+q^{|AC|}) \nonumber\\ &- (q^{|AD|} + q^{|BC|}) \bigg].
	\end{align}
	This can be used to estimate the mean value of $I^{(2)}(A:BC)$, which we argue in the main text probes the fidelity of the Hayden-Preskill teleportation protocol
	\begin{align}
		&\langle I^{(m)}(A:BC) \rangle_{U_t} \approx |AC| \log q \nonumber\\&- \log\left(\frac{q^{2N}(q^{|A| - |C|} + q^{|C| - |A|} - q^{-|AC|}) - q^{|AC|}}{q^{2N}-1}\right).
	\end{align}
	In the case of interest $|A| = |C|$, this becomes
	\begin{align}
	&= |AC| \log q - \log\left(\frac{q^{2N}(2-q^{-|AC|}) - q^{|AC|}}{q^{2N}-1}\right). 
	\label{eq:HaarRenyiMI}
	\end{align}
	The first term is the maximum value for the R{\'e}nyi mutual information. The second term, describing deviations from the maximum value, remains order one when one takes $|N| \rightarrow \infty$ while keeping $|A| = |C|$ fixed. This is consistent with the expectation that information about the initial state of $A$ can be recovered even if one only has access to a vanishing fraction of outputs $C$ (this corresponds to the amount of Hawking radiation in the Hayden-Preskill protocol \cite{Hayden2007}).   Evaluating \eqref{eq:HaarRenyiMI} for the case $N = 5$, $q = 2$, $|A| = |C| = 1$ (the parameters used for the data plotted in Fig.~\ref{fig:Mutinf5}), we find $I^{(2)}(A:BC) \approx 1.1945\ldots$.
	
	\section{Proof of Eq.~\eqref{eq:KLocal} \label{app:KLocal}}
	Here we prove the relationship between the quantities $D_k^C(t)$, which measure the $k$-locality of time-evolved operators that initially have support in $C$, and the operator purities $\Tr[\rho^{AC}(t)^2]$. Firstly, trace preservation implies that $\mathcal{N}^\dagger[\mathbb{I}] = \mathbb{I}$, which in turn gives $c^{I \nu}(t) = \delta_{\nu, I}$, where $I$ labels the identity Pauli string. Thus, for $k \geq 1$, the restriction $\mu \neq I$ in the sum on first line of \eqref{eq:KLocal} can be removed. Then, we consider the sum of operator purities over all subsets of qubits $A$ of fixed size $|A| = r$
	\begin{align}
	E^C_r(t) &\coloneqq \frac{2^{|C|+r}}{2^{|C|}-1} \sum_{A \subseteq S; |A| = r} \Tr[\rho^{AC}(t)^2] \\
	&= \frac{1}{2^{|C|}-1} \sum_{k = 0}^r {N - k \choose r - k} \sum_{\mu \in C}  \sum_{\nu : |\pstring^\nu| = k} |c^{\mu \nu}(t)|^2 \\
	&= \sum_{k = 0}^r {N - k \choose N-r} D_{k}^C
	\end{align}
	where for convenience we alter the definition of $D_k^C$ for $k = 0$ to be $D_0^C = \Tr[\rho^C(t)^2]/(2^{|C|} - 1)$, which differs from the expression \eqref{eq:KLocal} in the inclusion of the term $\mu = I$. The above follows from counting the number of subregions $A$ that support a Pauli string that acts non-trivially on $k$ qubits. This establishes a linear relationship between the sums $E_r^C(t)$ and the quantities of interest $D_k^C$, which can be inverted. The inverse of the lower triangular matrix $[L]_{r k} = {N - k \choose N-r}$ ($r \geq k$) is simply given by $[L^{-1}]_{k r} = (-1)^{k+r} {N - r \choose N-k}$ ($k \geq r$); this can be proved using the relation $\sum_{m = i}^j (-1)^{j+m} {j \choose m} {m \choose i} = \delta_{ij}$. This gives $D_k^C = \sum_{r = 0}^k (-1)^{r+k} {N - r \choose N-k} E_r^C(t)$, which can be easily manipulated to give Eq.~\eqref{eq:KLocal}.

	\section{Justification of Eq.~\eqref{eq:RhoEst} \label{app:RhoEst}}

	\begin{figure}
		\includegraphics[width=246pt]{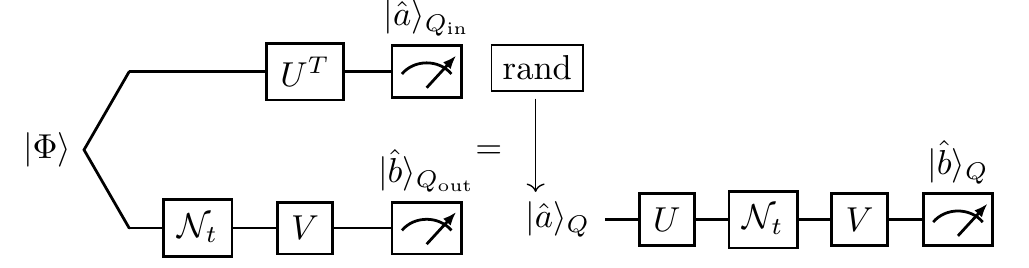}
		\caption{Left: Conventional shadow tomography on the operator state $\rho_{\rm op}(t) = (\text{id}_{\rm in} \otimes \mathcal{N}_t)[\Phi]$. The distribution of unitaries $U$, $V$ and measurement outcomes $\hat{a}$, $\hat{b}$ are the same as that of a hybrid classical-quantum process (right), where $\hat{a}$ are sampled from a uniform distribution, and then used as the input for a quantum circuit.}
		\label{fig:Doubled}
	\end{figure}
	
	In this section, we prove that the quantity \eqref{eq:RhoEst} is indeed an unbiased estimator of the operator-state $\rho_{\rm op}(t)$,
	i.e.~$\mathbb{E}[\hat{\rho}_{\rm op}(t)] = \rho_{\rm op}(t)$, where the expectation value is taken over the joint distribution of unitaries $U_j$, $V_j$, and outcomes $\hat{b}_j$. This can be done relatively straightforwardly using the graphical equation shown in Fig.~\ref{fig:Doubled}. First, suppose that one could explicitly construct $\rho_{\rm op}(t)$ in the experiment; then one could perform conventional shadow tomography, where unitaries $U$ and $V$ are applied to  $Q_{\rm in}$ and $Q_{\rm out}$, respectively, with outcomes $\hat{a}, \hat{b} \in \{0, 1\}^{\times N}$. This is shown on the left hand side of Fig.~\ref{fig:Doubled}. Using the property of the maximally mixed state $(O_{\rm in}^T \otimes \mathbb{I}_{\rm out})\ket{\Phi} = (\mathbb{I}_{\rm in} \otimes O_{\rm out})\ket{\Phi}$, one can push the unitary $U = \bigotimes_j U_j$ acting on $Q_{\rm in}$ onto the other half of the doubled system. This makes it clear that the distribution of measurements $\{\hat{a}\}$ on the input qubits is uniform over $\{0, 1\}^{\times N}$. Thus, we can sample $\hat{a}$ using a classical computer, and use it as the input to a circuit that only requires a single copy of the system (right hand side of Fig.~\ref{fig:Doubled}). The joint distribution of $\hat{a}$, $\hat{b}$, $U$, $V$ will be exactly the same as that of state shadow tomography on $\rho_{\rm op}(t)$, which allows us to construct an unbiased estimator of $\rho_{\rm op}(t)$ in the usual way \cite{Huang2020}.

	Finally, we note that the variables $\hat{a}$, $U$ only appear in the combination $U\ket{\hat{a}}$ in both the circuit and the shadow tomography estimator of the density matrix. Thus, we need only ensure that the ensemble of inputs to the channel $\mathcal{N}_t$ has the correct distribution. In our case $U$ is distributed uniformly over products of single-qubit Clifford operations; we can therefore replace $U\ket{\hat{a}}$ with $U\ket{0^{\otimes N}}$ without modifying the appropriate distribution. This justifies the form of Eq.~\eqref{eq:RhoEst}.
	
	\section{Results for $N = 7$ qubits \label{app:N7}}
	
	In this Appendix, we describe a circuit model of dynamics that uses all 7 qubits of the quantum device \texttt{ibm\_lagos}, and present results obtained from the shadow protocol. 
	
	\begin{figure}
		\centering
		\includegraphics[width=246pt]{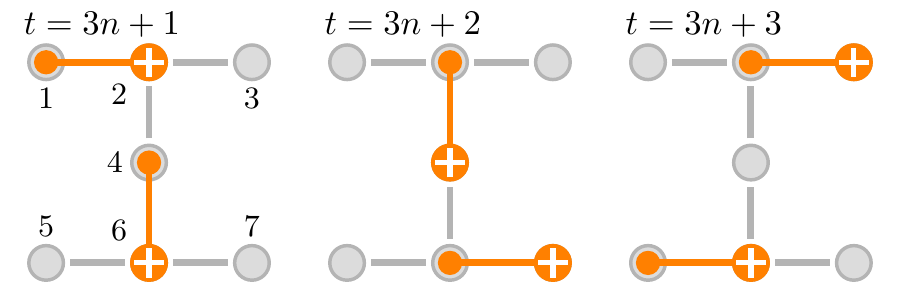}
		\caption{Configuration of the CNOT layers in the model of chaotic dynamics that uses all 7 qubits of \texttt{ibm\_lagos}. The pattern of CNOTs repeats every 3 timesteps. Small orange circles denote the control qubits, and large orange circles with a plus are the target qubits. Note that the numbering of the qubits (indicated in the leftmost panel) differs from that used in the main text.}
		\label{fig:N7Circuit}
	\end{figure}
	
	\begin{figure}
		\centering
		\includegraphics[width=246pt]{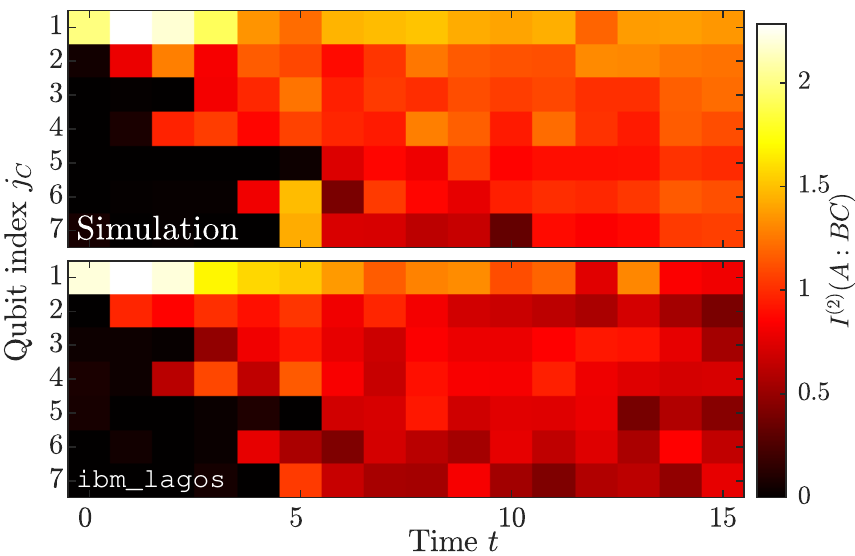}
		\caption{Color plot of the second R{\'e}nyi mutual information $I^{(2)}(A:BC)$ for the circuit model of dynamics described in Appendix \ref{app:N7}, using all 7 qubits of \texttt{ibm\_lagos}. We fix $A = \{1\}$ [the bottom right qubit in Fig.~\ref{fig:RenyiMI}(a)], and $C = \{j_C\}$, where $j_C$ is varied. Top panel: results obtained from noiseless classical simulations of the time evolution $\mathcal{N}_t$ and the shadow protocol. By averaging the deviation of these data from the exact value of $I^{(2)}(A:BC)$ for the circuit in question, we obtain an estimate of the statistical fluctuations due to the shadow protocol of $0.07$. Bottom panel: data obtained from \texttt{ibm\_lagos}.}
		\label{fig:N7Renyi}
	\end{figure}

	To generate chaotic dynamics, we use a circuit design made up of the same gates as the setup presented in the main text [Fig.~\ref{fig:RenyiMI}(b)], namely CNOTs and single-qubit gates independently sampled from the discrete set $\{W_c : c = 1, \ldots, 4\}$. As before, each timestep is made up of a layer of single-qubit unitaries acting on all qubits followed by a layer of CNOTs. The arrangement of CNOTs changes each timestep, repeating itself after a period of $3$ steps, as illustrated in Fig.~\ref{fig:N7Circuit}. This ensures that entanglement can generated between any two qubits after a sufficient amount of time. 
	
	After running the shadow tomography protocol with the same parameters as before ($N_U = 900$, $N_M = 8192$), the R{\'e}nyi mutual information $I^{(2)}(A:BC)$ was computed, where we set $A = \{1\}$, the top left qubit in Fig.~\ref{fig:N7Circuit}. We also generate a set of shadow data by simulating the full circuit without noise on a classical computer, for comparison. The results are presented in Fig.~\ref{fig:N7Renyi}. Initially, correlations are only present for $j_C = 1$, whereas at later times these correlations are distributed across the entire system, thus confirming that information has been scrambled. As before, the values from \texttt{ibm\_lagos} at later times are systematically below those from classical simulations, due to noisy processes that disturb the propagation of information.
	
	The region $ABC$ involves more qubits than that used for the $N = 5$ setup described in the main text, and so we expect to incur larger statistical errors when computing the operator-space R{\'e}nyi entropies, and in turn $I^{(2)}(A:BC)$. The size of these errors can be estimated by looking at the deviation of the values from noiseless classical simulations of the shadow protocol, compared with the exact values of the mutual information. Averaging across all times $t$ and choices of $j_C$, we find a mean relative error in the value of $\Tr[\rho_{ABC}^2]$ of $0.05$, and an absolute error in $I^{(2)}(A:BC)$ of $0.07$. Evidently, even for regions as large as $|ABC| = 8$, it is possible to estimate R{\'e}nyi entropies and quantities derived thereof to a good accuracy using a reasonable number of shots.

	\bibliography{shadow_opent.bib}
	\bibliographystyle{apsrev4-1}
	
	\newpage
	\afterpage{\blankpage}
	
	\newpage
	
	\setcounter{equation}{0}
	\setcounter{figure}{0}
	\setcounter{table}{0}
	\setcounter{page}{1}
	
	\renewcommand{\theequation}{S\arabic{equation}}
	\renewcommand{\thefigure}{S\arabic{figure}}
	
	\begin{widetext}
	\begin{center}
		{\fontsize{12}{12}\selectfont
			\textbf{Supplemental Material for ``\papertitle''\\[5mm]}}
\		{\normalsize \authorlist\\[1mm]}
		
	\end{center}
\normalsize
\end{widetext}
	\

	\subsection*{Repeating random unitaries in shadow tomography}
	
	While shadow tomography is ideally performed using different measurement bases for each shot, in some platforms it is possible to achieve a higher total shot count by running each circuit multiple times. This is the approach we use to obtain the data used to generate the quantities plotted in the main text. The circuits are designed as follows: First, starting from an initial state $\ket{0^{\otimes N}}$, a Hadamard gate is applied to each qubit, followed by measurements of all qubits in the computational basis. This generates a random initial computational state $\ket{\Psi_{\hat{a}}} = \bigotimes_j \ket{\hat{a}_j}$, where $\hat{a}_j \in \{0, 1\}$. This way, both $\hat{a}_j$ and $\hat{b}_j$ are random variables that are sampled independently for different shots of the same circuit, whereas $U_j$ and $V_j$ are fixed for a particular circuit. We can verify \textit{a posteriori} that the distribution of $\hat{a}_j$ is uniform. The rest of the shadow tomography protocol proceeds as usual (Fig.~\ref{fig:Protocol}), with $\ket{\Psi_{\hat{a}}}$ in place of the ordinary initial state $\ket{0^{\otimes N}}$. The basis rotations $U_j$, $V_j$ are fixed for a particular circuit. A total of $M_{\rm U}$ circuits are generated, and each is run $M_{\rm S}$ times.
	
	This scenario where circuits are repeated multiple times is closer to the protocol for measuring R{\'e}nyi entropies proposed by Elben \textit{et al.}~\cite{Elben2018}, which was implemented in Ref.~\cite{Brydges2019}. Interestingly, it is possible to understand both this method and the usual shadow tomography process using the same formalism, as we now explain. We will focus solely on the second R{\'e}nyi entropy, which was the main quantity considered in Refs.~\cite{Elben2018, Brydges2019}. Additionally, for now we drop the distinction between state and channel shadow tomography, simply referring to a state $\rho$ with a total of $N$ qubits.
	
	The purity $P = \Tr[\rho^2]$ is quadratic in the density matrix. Thus, unbiased estimators of $P$ should be constructed using correlations between \textit{pairs} of different experiments. Let us pick such a pair from the total of $M = M_{\rm S}M_{\rm U}$ experiments. For $M_{\rm S}^2 M_{\rm U}(M_{\rm U} - 1)/2$ of these pairs (which we call type I), the two experiments will correspond to independently generated circuits, i.e.~$U$ and $V$ will be different, while the remaining $M_{\rm U}M_{\rm S}(M_{\rm S} - 1)/2$ pairs (type II) will correspond to two different shots of the same circuit.
	
	For a given pair of either type, one can describe the probability distribution of possible outcomes using a positive operator-valued measure (POVM) --- a collection of positive operators $\{F_\mu\}$ acting on a doubled Hilbert space (each factor representing one of the two experiments), satisfying $\sum_\mu F_\mu = \mathbb{I}$. The joint index $\mu$ enumerates the possible data that could arise from the pair of experiments; namely, the classical bit strings $\hat{b}_j^{(r)}$ (where $r = 1,2$ labels the two runs), and the unitaries $V^{(r)}$. The probability of obtaining the joint outcome $\mu$ is then given by $\Tr[F_\mu (\rho\otimes \rho)]$.
	
	At this point, it becomes helpful to view operators $O$ over the doubled Hilbert space $\mathcal{H}^{\otimes 2}$ as vectors $|O\rrangle$ in a linear space $\mathcal{B}(\mathcal{H}^{\otimes 2})$, endowed with the Hilbert-Schmidt inner product $\llangle X | Y \rrangle \coloneqq d^{-2}\Tr[X^\dagger Y]$, where $d = |\mathcal{H}|$ is the Hilbert space dimension. One can then express the POVM as a linear map $\mathcal{E} = d^2 \sum_\mu |\mu) \llangle F_\mu|$ taking doubled quantum states $|\sigma\rrangle \in \mathcal{B}(\mathcal{H}^{\otimes 2})$ to classical probability distributions. Here, we treat the space of probability distributions as a linear space itself with basis vectors $|\mu )$, such that a collection of probabilities $\{p_\mu\}$ becomes a vector $|p) = \sum_\mu p_\mu |\mu)$. In this language, the distribution of outcomes $|p) = \mathcal{E}|\rho \otimes \rho\rrangle$. The POVM property of $\{F_\mu\}$ translates to $\mathcal{E}$ being completely positive and trace-preserving (CPTP).
	
	For type I pairs, the unitaries are sampled independently, so we have
	\begin{widetext}
	\begin{align}
	F_\mu &= q(V^{(1)}) q(V^{(2)}) \Big[[V^{(1)}]^\dagger\ket{\hat{b}^{(1)}}\bra{\hat{b}^{(1)}}V^{(1)} \otimes [V^{(2)}]^\dagger\ket{\hat{b}^{(2)}}\bra{\hat{b}^{(2)}}V^{(2)}\Big], & \text{type I}
	\end{align}
	where $q(V)$ is the classical probability distribution for selecting the unitary $V$. For type II pairs, the unitaries are the same for the two experiments, so
	\begin{align}
	F_\mu &= \delta_{V^{(1)} = V^{(2)}} q(V^{(1)}) \Big[ [V^{(1)}]^\dagger\ket{\hat{b}^{(1)}}\bra{\hat{b}^{(1)}}V^{(1)} \otimes [V^{(1)}]^\dagger\ket{\hat{b}^{(2)}}\bra{\hat{b}^{(2)}}V^{(1)}\Big]. & \text{type II}
	\end{align}
	\end{widetext}
	This defines two distinct channels $\mathcal{E}_{\rm I}$, $\mathcal{E}_{\rm II}$ as described above.
	
	Now, an estimator $\hat{P}$ for $P$ can be expressed as a map taking an outcome $\mu$ and returning a scalar: $\hat{P} : \mu \mapsto w_\mu$. We can express this as a dual vector $(P| = \sum_\mu w_\mu (\mu|$, such that the expectation value of the estimator is $\mathbb{E}[\hat{P}] = (P|p) = (P|\mathcal{E}|\rho \otimes \rho\rrangle$. Suppose that $\mathcal{E}$ has an inverse $\mathcal{E}^{-1}$ (as is the case if the POVM is informationally complete \cite{Busch1991}). Then, since $P = \Tr[\Pi \rho \otimes \rho]$, where $\Pi(\ket{\phi} \otimes \ket{\psi}) = \ket{\psi} \otimes \ket{\phi}$ is the swap operator, we should choose $|P) =  \mathcal{E}^{-1}| \Pi \rrangle$, whence $\mathbb{E}[\hat{P}] = \llangle \Pi | \mathcal{E}^{-1} \mathcal{E}| \rho \otimes \rho\rrangle = P$, as desired. Since $\mathcal{E}_{\rm I}$ is informationally complete \cite{Huang2020}, we can compute this estimator, and we recover the expression given in Ref.~\cite{Huang2020} for the estimator of the purity [see Eq.~\eqref{eq:RenyiEst} with $m = 2$].
	
	However, even if $\mathcal{E}$ does not have an inverse, it may still be possible to define a pseudoinverse $\mathcal{E}^+$ on the space spanned by $\Pi$ (satisfying $\mathcal{E}^+\mathcal{E} = \mathcal{P}$, where $\mathcal{P}$ is a projector in operator space satisfying $\mathcal{P}|\Pi\rrangle = |\Pi\rrangle$). In this case $|P) = \mathcal{E}^+|\Pi\rrangle$ defines an unbiased estimator of the purity. This is indeed the case for $\mathcal{E}_{\rm II}$. A straightforward (though tedious) calculation confirms that the resulting expression for $|P)$ corresponds to the expression provided in Ref.~\cite{Elben2018, Brydges2019} for the purity.
	
	In conclusion, from the combination of $M_{\rm S} M_{\rm U}$ sets of experimental data, one can construct estimators of the purity for each pair of experiments. The expression for each estimator depends on whether the pair corresponds to the same or independently generated circuits. The method used in Refs.~\cite{Elben2018, Brydges2019} makes use of the type II estimators only. In contrast, the classical shadow protocol uses the limit $M_{\rm S} = 1$, such that only type I estimators remain.
	
	In our case, we have access to both types of estimator. In principle, a minimum-variance estimator could be constructed as an optimal linear combination of all type I and type II estimators. Here, for ease of implementation, we use the type I estimators only. This is equivalent to constructing shot-averaged density matrices $\hat{\rho}^{(r_{\rm U})}_{\rm avg} = M_{\rm S}^{-1} \sum_{r_{\rm S} = 1}^{M_{\rm S}} \hat{\rho}^{(r_{\rm U}, r_{\rm S})}$ (where $\hat{\rho}^{(r_{\rm U}, r_{\rm S})}$ is Eq.~\eqref{eq:RhoEst} for circuit index $r_{\rm U}$ and shot index $r_{\rm S}$) for each measurement basis, and computing
	\begin{align}
	\hat{P} = {M_{\rm U} \choose 2}^{-1} \sum_{r_1 = 1}^{M_{\rm U}} \sum_{r_2 = r_1 + 1}^{M_{\rm U}} \Tr[\hat{\rho}_{\rm avg}^{(r_1)} \hat{\rho}_{\rm avg}^{(r_2)} ].
	\label{eq:PurityPreAverage}
	\end{align}
	We leave the problem of determining the optimum combination of estimators to future work.
	
	The statistical errors coming from this process are sub-optimal for a fixed measurement budget $M = M_{\rm S} M_{\rm U}$. Nevertheless, increasing $M_{\rm S}$ for fixed $M_{\rm U}$ (which can be done efficiently on the IBM system that we use) can decrease the errors, particular for highly mixed states. This is because the shot-averaged density matrices $\hat{\rho}^{(r_{\rm U})}_{\rm avg}$ typically have a narrower spectrum than the individual objects of the form \eqref{eq:RhoEst}. Thus, the individual terms in the double sum in Eq.~\eqref{eq:PurityPreAverage} will be smaller, and the full average will converge more quickly. Note, however, that taking $M_{\rm S} \rightarrow \infty$ for fixed $M_{\rm U}$ does not reduce the error to zero.

	\subsection*{Error analysis}
	
	In this section, we compute the variance of the estimator of moments of the reduced density matrix, which in turn determines how many experimental runs $N$ are needed to predict the R{\'e}nyi entropy to a desired accuracy. Specifically, we consider
	\begin{align}
	\hat{\theta}_m \coloneqq \frac{(M-m)!}{M!} \sum_{r_1 \neq \cdots \neq r_m} \prod_{j \in AC} \Tr[\hat{\rho}_j^{(r_1)} \cdots  \hat{\rho}_j^{(r_m)}],
	\label{eq:RenyiEst}
	\end{align}
	which is an estimator for $\theta_{m} \coloneqq \Tr[(\rho_{AC})^m]$. Here, $r_i \in \{1, \ldots, M\}$ indexes the different experimental runs, and $\hat{\rho}_j^{(r)}$ is one of the factors in Eq.~\eqref{eq:RhoEst} corresponding to qubit $j$ (input or output), and run $r$. Since these bounds are expected to be indicative of the qualitative form of scaling, rather than being quantitatively tight \cite{Huang2020}, we will consider the ideal shadow tomography measurement allocation $M_{\rm S} = 1$, with the expectation that similar behaviour should be expected for $M_{\rm S} > 1$, at least in the regime $M_{\rm U} \gg M_{\rm S}$.
	
	Being an example of a $U$-statistic, $\Var[\hat{\theta}_m]$ can be reduced to standard formulae as outlined in, e.g.~Ref.~\cite{Ferguson2003}. We briefly summarise these derivations before evaluating the variance for our specific problem.
	
	We first find it useful to re-express the estimator as
	\begin{align}
	\hat{\theta}_m = \begin{pmatrix}
	M \\ m
	\end{pmatrix}^{-1} \sum_{r_1 <  \cdots < r_m} h(\hat{\rho}^{(r_1)}_{AC}, \ldots, \hat{\rho}^{(r_m)}_{AC}).
	\end{align}
	Here, $\hat{\rho}_{AC}^{(r)} \coloneqq \bigotimes_{j \in AC} \hat{\rho}_j^{(r)}$, and we have defined a function of $m$ density operators
	\begin{widetext}
	\begin{align}
	h(\rho_1, \ldots, \rho_m) &= \frac{1}{m!}\sum_{\sigma \in \Sigma_m} \Big\langle \rho_1, \ldots, \rho_m \Big\rangle_\sigma & \text{where } \Big\langle \rho_1, \ldots, \rho_m \Big\rangle_\sigma \coloneqq \Tr[\rho_{\sigma(1)} \cdots \rho_{\sigma(m)}],
	\label{eq:TracePerm}
	\end{align}
	where the sum is over all permutations of $m$ elements. Now, by definition we have
	\begin{align}
	\Var[\hat{\theta}_m] = \begin{pmatrix}
	M \\ m
	\end{pmatrix}^{-2} \sum_{r_1 \neq  \cdots \neq r_m} \sum_{s_1 \neq  \cdots \neq s_m} \Cov\left(h(\hat{\rho}^{(r_1)}_{AC}, \ldots, \hat{\rho}^{(r_m)}_{AC}),\; h(\hat{\rho}^{(s_1)}_{AC}, \ldots, \hat{\rho}^{(s_m)}_{AC}) \right).
	\label{eq:UstatVarTheta}
	\end{align}
	The covariance in the above can be expressed with the help of the following quantities, defined for $c = 0, 1, \ldots, m$
	\begin{align}
	h_c(\rho_1, \ldots, \rho_c) \coloneqq \mathbb{E}_{\hat{\rho}_{c+1},  \ldots, \hat{\rho}_{m}}[h(\rho_1, \ldots, \rho_c, \hat{\rho}_{c+1},\ldots, \hat{\rho}_{m})] = h(\rho_1, \ldots, \rho_c, \underbrace{\rho_{AC}, \ldots, \rho_{AC}}_{m-c\text{ copies}})
	\label{eq:UstatHc}
	\end{align}
	which gives the expectation value of $h$ over the random variables $\hat{\rho}_{c+1}, \ldots, \hat{\rho}_{m}$ [drawn independently from the same distribution as each $\hat{\rho}^{(r)}_{AC}$], with the density matrices $\rho_1, \ldots, \rho_c$ fixed. We have used the fact that $h$ is linear in each of its arguments. If we then use the random variables $\hat{\rho}_1, \ldots, \hat{\rho}_c$ as arguments to the function \eqref{eq:UstatHc}, then we evidently have $\mathbb{E}_{\hat{\rho}_1, \ldots, \hat{\rho}_c}[ h_c(\hat{\rho}_1, \ldots, \hat{\rho}_c)] = \theta_m$, where $\theta_m \coloneqq \mathbb{E}[\hat{\theta}_m] = \Tr[(\rho_{AC})^m]$, and we define the variance $\sigma_c^2 \coloneqq \Var_{\hat{\rho}_1, \ldots, \hat{\rho}_c}[h_c(\hat{\rho}_1, \ldots, \hat{\rho}_c)]$. One can then show that \cite{Ferguson2003}
	\begin{align}
	\Cov\left(h(\hat{\rho}^{(r_1)}_{AC}, \ldots, \hat{\rho}^{(r_m)}_{AC}),\; h(\hat{\rho}^{(s_1)}_{AC}, \ldots, \hat{\rho}^{(s_m)}_{AC}) \right) = \sigma_c^2
	\end{align}
	where $c$ is the number of indices in $\{s_i\}$ that are also present in $\{r_i\}$. Eq.~\eqref{eq:UstatVarTheta} can then be evaluated in terms of combinatoric factors by counting how many sets of indices in the sums over $\{r_i\}$, $\{s_i\}$ have exactly $c$ indices in common. As argued in Ref.~\cite{Ferguson2003}, one finds
	\begin{align}
	\Var[\hat{\theta}_m] =\begin{pmatrix}
	M \\ m
	\end{pmatrix}^{-1} \sum_{c=1}^m \begin{pmatrix}
	m \\ c
	\end{pmatrix}\begin{pmatrix}
	M-m \\ m-c
	\end{pmatrix} \sigma_c^2
	\label{eq:TotVarianceSum}
	\end{align}
	(Note that $\sigma_0^2 = 0$ trivially.) The above formula applies to $U$-statistics in general. Now it remains to determine $\sigma_c^2$ for our specific problem. This is a challenging task to do exactly; however it is possible to derive sensible upper bounds, such as those given in Ref.~\cite{Huang2020}. We will rely heavily on a particular inequality: For any operator $O$ that acts non-trivially on $k$ qubits (i.e.~$O = \mathbb{I}_{2^{n-k}} \otimes \tilde{O}$), and for any underlying state $\rho$, the fluctuations of expectation values between different snapshots can be bounded by
	\begin{align}
	\Var_{\hat{\rho}}\left[\big|\!\Tr[O \hat{\rho}]\big|\right] \leq \mathbb{E}_{\hat{\rho}}\left[\big|\!\Tr[O \hat{\rho}]\big|^2\right] \leq  3^k \Tr[\tilde{O}^\dagger \tilde{O}]  \|\tilde{\rho}\|_\infty
	\label{eq:QubitIneq}
	\end{align}
	where $\tilde{\rho}$ is the reduced density matrix of $\rho$ on the region where $O$ acts non-trivially, and $\| X\|_{\infty} \coloneqq \max {\rm eig}\, X$ is the spectral norm. The proof of \eqref{eq:QubitIneq} is given at the end of this section. Note that an alternative bound for the same quantity was given in Ref.~\cite{Huang2020}:
	\begin{align}
	\Var_{\hat{\rho}}\left[\big|\!\Tr[O \hat{\rho}]\big|\right] &\leq 2^k \Tr[O^\dagger O]
	\label{eq:HuangIneq}
	\end{align}
	Eq.~\eqref{eq:QubitIneq} is an improvement on the above when the min-entropy $S^{(\infty)}_{AC} \coloneqq \min_i (-\log p_i) = -\log \| \rho_{AC} \|_\infty$ (where $p_i$ are the eigenvalues of $\rho_{AC}$) exceeds $|AC| \log(3/2)$, which is to be expected for highly mixed states.
	
	With Eq.~\eqref{eq:QubitIneq} in hand, it is instructive to first consider the quantity $\sigma_c^2$ for $c = 1$, which can be bounded as
	\begin{align}
	\sigma_1^2 &\leq \mathbb{E}_{\hat{\rho}}\left[\Tr[(\rho_{AC})^{m-1} \hat{\rho}]^2\right] \nonumber\\
	&\leq 3^{|AC|} \Tr[(\rho_{AC})^{2m-2}] \|\rho\|_\infty^2 \nonumber\\
	&=   \exp\Big(|AC|\log 3 - 2(m-1)S^{(2m-2)}_{AC} - S_{AC}^{(\infty)}\Big).
	\label{eq:VarFirst}
	\end{align}
	For $c = 2$, we group permutations in the sum in Eq.~\eqref{eq:TracePerm} together, giving
	\begin{align}
	\sigma_2^2 = \Var_{\hat{\rho}_1, \hat{\rho}_2}\left[ \frac{m\times (m-2)!}{m!}  \sum_{u = 0}^{m-2} \Tr[\hat{\rho}_1 (\rho_{AC})^u \hat{\rho}_2 (\rho_{AC})^{m-2-u}] \right].
	\label{eq:Sigma2}
	\end{align}
	The above can be re-expressed using the identity $\Tr[ABCD] = \Tr[(B \otimes D) \Pi_{\leftarrow} (A \otimes C)]$, where $\Pi_{\leftarrow}$ is a swap operator acting between two copies of the Hilbert space, i.e.~$\Pi_{\leftarrow} (\ket{\phi_1} \otimes \ket{\phi_2}) = \ket{\phi_2} \otimes \ket{\phi_1}$ for all wavefunctions $\ket{\phi_1}$, $\ket{\phi_2}$. The trace in Eq.~\eqref{eq:Sigma2} then becomes $\Tr[(\rho_{AC}^u \otimes \rho_{AC}^{m-2-u}) \Pi_{\leftarrow} (\hat{\rho}_1 \otimes \hat{\rho}_2)]$. Since $\hat{\rho}_1 \otimes \hat{\rho}_2$ is a classical snapshot of the state $\rho \otimes \rho$, we can substitute the $2k$-qubit operator $O = (\sum_{u=0}^{m-2} \rho_{AC}^u \otimes \rho_{AC}^{m-2-u})\Pi_{\rightarrow}$ into Eq.~\eqref{eq:QubitIneq}, after the replacement $k \rightarrow 2k$. This gives
	\begin{align}
	\sigma_2^2 &\leq (m-1)^{-2} \times 3^{2|AC|} \|\rho \otimes \rho\|_\infty \Tr\left[ \left|\left[ \sum_{u=0}^{m-2} (\rho_{AC})^u \otimes (\rho_{AC})^{m-2-u} \right] \Pi_{\rightarrow}\right|^2 \right] \nonumber\\
	&= (m-1)^{-2} \times 3^{2|AC|} \|\rho\|_\infty^2 \left[(m-1)\Tr[(\rho_{AC})^{m-2}]^2 + \sum_{s = 0}^{m-3} 2(s+1)\Tr[(\rho_{AC})^s] \Tr[(\rho_{AC})^{2m - 4 - s}]\right] \nonumber\\
	&\leq 3^{2|AC|} \|\rho\|_\infty^2 \times 2^{|AC|} \Tr[(\rho_{AC})^{2m-4}] \nonumber\\ &= (m!)^2\exp\Big( |AC|\log[18] - 2(m-2) S_{AC}^{(2m-4)} - 2 S_{AC}^{(\infty)} \Big)
	\end{align}
	where in the second line we use $\Pi_{\rightarrow}^\dagger \Pi_{\rightarrow}^{\vphantom{\dagger}} = \mathbb{I}$, and re-expressed the two sums over $u$ coming from the two factors in $\Tr[O^\dagger O]$ as a single sum over $s$. In the third line we use $\Tr[(\rho_{AC})^s]\Tr[(\rho_{AC})^{2m-4-s}] \leq d_{AC} \Tr[(\rho_{AC})^{2m-4}]$. (To see this, write $\Tr[\rho^{b+a}]\Tr[\rho^{b-a}] = \sum_{jk} \lambda_j^b \lambda_k^b f_a(\lambda_k/\lambda_j)$, where $\{\lambda_j\}$ are the eigenvalues of $\rho$, and $f_a(x) \coloneqq (x^a + x^{-a})/2$; then note that $f_a(x)$ is a non-decreasing function of $a$ for $x, a > 0$, and so is maximized when $a = b$.)
	
	Generalising the above approach to include $c > 2$, we find
	\begin{align}
	\sigma_c^2 = \Var_{\hat{\rho}_1, \ldots, \hat{\rho}_c}\left[ \frac{m \times (m-c)!}{m!} \sum_{u_1 = 0}^{m-c} \sum_{u_2 = 0}^{m-c-u_1} \cdots \sum_{u_{c-1} = 0}^{m - c - \sum_{k=1}^{c-2} u_k} \Tr[\hat{\rho}_1 (\rho_{AC})^{u_1} \cdots \hat{\rho}_{c-1} (\rho_{AC})^{u_{c-1}} \hat{\rho}_c (\rho_{AC})^{m - c - \sum_k u_k}] \right]
	\end{align}
	The trace in the above can be written as $\Tr[(\rho_{AC}^{u_1} \otimes \cdots \otimes \rho_{AC}^{u_{c-1}} \otimes \rho_{AC}^{m-c-\sum_k u_k}) \Pi_{\leftarrow} (\hat{\rho}_1 \otimes \cdots \otimes \hat{\rho}_c)]$, where now $\Pi_{\leftarrow}$ is a cyclic permutation operator acting on $c$ copies $\Pi_{\leftarrow}(\ket{\phi_1} \otimes \cdots \otimes \ket{\phi_c}) = \ket{\phi_2} \otimes \cdots \otimes \ket{\phi_c} \otimes \ket{\phi_1}$. We thus have
	\begin{align}
	\sigma_c^2 &= \left(\frac{(m-c)!}{(m-1)!}\right)^2 \times 3^{c|AC|} \|\rho_{AC}\|_\infty^c  \sum_{\{r_i\}, \{s_i\}}  \Tr[\rho_{AC}^{r_1 + s_1}] \cdots \Tr[\rho_{AC}^{r_c + s_c}]
	\nonumber\\
	&\leq [(c-1)!]^{-2}\times 3^{c|AC|} 2^{(c-1)|AC|} \|\rho_{AC}\|_\infty^c \Tr[\rho^{2m-2c}] \nonumber\\
	&= [(c-1)!]^{-2}\times \exp\Big( |AC|\log[2^{c-1}3^c] - (2m-2c)S_{AC}^{(2m-2c)} - cS^{(\infty)}_{AC}  \Big)
	\end{align}
	where the sums in the first line are restricted to $\sum_i r_i = \sum_i s_i = m - c$. We use the fact that the summand is always less than or equal to $\Tr[\mathbbm{I}]^{c-1} \Tr[\rho^{\sum_j r_j + s_j}]$, and that there are ${m-1 \choose c-1}^2$ terms in total. Evidently, if we had used the bound \eqref{eq:HuangIneq} instead of \eqref{eq:QubitIneq}, we would find an alternative bound
	\begin{align}
	\sigma_c^2 \leq [(c-1)!]^{-2}\times \exp\Big( (2c-1)|AC| - (2m-2c)S_{AC}^{(2m-2c)} \Big)
	\end{align}
	Finally, putting everything together, and using ${M\choose m}^{-1} {m \choose c} {M-m \choose m-c} \leq N^{-c}[m!/(m-c)!]^2 / c!$, we find
	\begin{align}
	\Var[\hat{\theta}_m] \leq   \sum_{c=1}^m {m \choose c}^2  \frac{c^2}{c!} \times \frac{1}{M^c} \exp\Big( |AC|\log[2^{c-1}3^c] - (2m-2c)S_{AC}^{(2m-2c)} - cS^{(\infty)}_{AC}  \Big)
	\end{align}
	In the $M \rightarrow \infty$ limit, the right hand side of the above will be dominated  by the $c = 1$ term, which simplifies using \eqref{eq:VarFirst}, and scales as $M^{-1/2}$ as expected.

	\newcommand{\bp}{\mathbf{p}}
	\newcommand{\bq}{\mathbf{q}}
	\newcommand{\bs}{\mathbf{s}}
	\newcommand{\sop}[1]{{\underline{\underline{#1}}}}
	
	\textit{Proof of Eq.~\eqref{eq:QubitIneq}.---} Our derivation follows that of Proposition S3 in Ref.~\cite{Huang2020}, with the difference that we do not maximise over all underlying states $\rho$. We can expand $\tilde{O}$ in a basis of Pauli operators acting on $k$ qubits $\tilde{O} = \sum_{\bp} a_\bp P_\bp$, where $\bp \in \{I, X, Y, Z\}^k$, and $P_\bp = \sigma_{p_1} \otimes \cdots \otimes \sigma_{p_k}$. The Pauli operators are orthonormal under the Hilbert-Schmidt inner product $\llangle P_\bp | P_\bq \rrangle \coloneqq 2^{-k} \Tr[P_\bp^\dagger P_\bq] = \delta_{\bp, \bq}$. By taking appropriate averages over the random unitaries, one can show that \cite{Huang2020}
	\begin{align}
	\mathbb{E}_{\hat{\rho}}\left[\big|\!\Tr[O \hat{\rho}]\big|^2\right] &= \sum_{\bp \bq} a_\bp^* a_\bq f(\bp, \bq) \Tr[\rho P_\bp P_\bq]
	\end{align}
	where $f(\bp, \bq) = \prod_{j=1}^k f_j(p_j, q_j)$, with the function $f_j(p_j,q_j)$ equal to 1 if $p_j = I$ or $q_j = I$; 3 if $p_j = q_j \neq I$; or 0 otherwise. Evidently, $f(\bp, \bq)$ is only non-zero if $\bp$ and $\bq$ can be obtained from the same vector $\bs \in \{X, Y, Z\}^k$ by replacing various elements with $I$. Because of this, the above can be written
	\begin{align}
	\mathbb{E}_{\hat{\rho}}\left[\big|\!\Tr[O \hat{\rho}]\big|^2\right] &= \sum_{\bs \in \{X, Y, Z\}^k} \sum_{\bp, \bq \rhd \bs} 3^{\sum_j |p_j||q_j|} \times \frac{1}{3^{\sum_j (1-|p_j|)(1-|q_j|)}} a_\bp^* a_\bq \Tr[\rho P_\bp P_\bq] \nonumber\\
	&= \frac{1}{3^k} \sum_{\bs \in \{X, Y, Z\}^k} \sum_{\bp, \bq \rhd \bs} 3^{|\bp| + |\bq|} a_\bp^* a_\bq \Tr[\rho P_\bp P_\bq]
	\end{align}
	Here we adopt the notation of \cite{Huang2020}, where $\bp \rhd \bs$ indicates that $\bp$ can be obtained from $\bs$ by setting a subset of elements to $I$. We define $|p_j| = 0$ if $p_j = I$, and $|p_j| = 1$ if $p_j \in \{X, Y, Z\}$; similarly $|\bp| = \sum_j |p_j|$ is the number of non-trivial Pauli operators in the string $\bp$. Note that the denominator in the first line is necessary to avoid over-counting.
	
	Now, we can define operators $\tilde{O}_\bs = \sum_{\bp \rhd \bs} a_\bp P_\bp$, which contain the components of $\tilde{O}$ within the subspace spanned by operators $\{P_\bp : \bp \rhd \bs\}$. We then have
	\begin{align}
	\mathbb{E}_{\hat{\rho}}\left[\big|\!\Tr[O \hat{\rho}]\big|^2\right] =\frac{2^k}{3^k} \sum_{\bs \in \{X, Y, Z\}^k} \llangle \tilde{O}_\bs | \mathbbm{f}\, \mathbbm{P}\, \mathbbm{f} | \tilde{O}_\bs\rrangle
	\end{align}
	where $\mathbbm{f}$ and $\mathbbm{P}$ are superoperators (i.e.~linear maps between operators), whose action on the Pauli basis is $\llangle P_\bp|\mathbbm{f}|P_\bq\rrangle = 3^{|\bp|} \delta_{\bp, \bq} $ and $\llangle P_\bp|\mathbbm{P}|P_\bq\rrangle = \Tr[\rho P_\bp^\dagger P_\bq]$. Matrix norms for superoperators can be defined in the usual way; in particular we consider the spectral norm $\|\mathbbm{P}\|_{\infty} \coloneqq \sup_{C :\, \llangle C |C \rrangle = 1} \llangle C | \mathbbm{P}|C\rrangle$. By the definition of this spectral norm, we have
	\begin{align}
	\mathbb{E}_{\hat{\rho}}\left[\big|\!\Tr[O \hat{\rho}]\big|^2\right] &\leq \frac{2^k}{3^k} \| \mathbbm{P}\|_{\infty} \sum_{\bs \in \{X, Y, Z\}^k} \llangle \tilde{O}_\bs | \mathbbm{f}^2 | \tilde{O}_\bs\rrangle  \nonumber\\
	&\leq \frac{2^k}{3^k} \| \mathbbm{P}\|_{\infty} \| \mathbbm{f}\|_{\infty} \sum_{\bs \in \{X, Y, Z\}^k}  \llangle \tilde{O}_\bs |\mathbbm{f}| \tilde{O}_\bs\rrangle  ,
	\end{align}
	Evidently, $\|\mathbbm{f}\|_{\infty} = 3^k$. Then, starting from $\|\mathbbm{P}\|_\infty = \sup_{C :\, \llangle C |C \rrangle = 1} 2^{-k} \Tr[C^\dagger C \rho]$, we expand $C^\dagger C = \sum_a p_a \ket{\phi_a}\bra{\phi_a}$, where $\ket{\phi_a}$ are normalized wavefunctions, and the coefficients satisfy $\sum_a p_a = \Tr[C^\dagger C] = 2^k \llangle C | C \rrangle = 2^k$. Thus, $\|\mathbbm{P}\|_\infty = 2^{-k}\sup_{p_a, \ket{\phi_a}:\, \sum_a p_a = 2^k}(p_a \braket{\phi_a|\rho|\phi_a}) \leq \|\rho\|_\infty$. Finally, following the arguments in Ref.~\cite{Huang2020}, we have
	\begin{align}
		\sum_{\bs \in \{X, Y, Z\}^k} \llangle \tilde{O}_\bs |\mathbbm{f}| \tilde{O}_\bs\rrangle = \sum_{\bs \in \{X, Y, Z\}^k} \sum_{\bp \rhd \bs} 3^{|\bp|} |a_\bp|^2 = 3^k \sum_{\bp} |a_\bp|^2 = \frac{3^k}{2^k} \Tr[\tilde{O}^\dagger \tilde{O}]
	\end{align}
	 Putting this all together, we arrive at Eq.~\eqref{eq:QubitIneq}.
	\end{widetext}
	
\end{document}